\documentclass[aps, prl, onecolumn, groupedaddress,
  superscriptaddress, shortbibliography, notitlepage]{revtex4-1}

\usepackage{graphicx} 
\usepackage{color} 
\usepackage{mathrsfs, amsmath}
\usepackage{amsfonts} 
\usepackage{amssymb}
\usepackage{upgreek}
\usepackage{dcolumn} 
\usepackage{comment}
\usepackage[justification=centerlast]{caption}
\usepackage[draft]{todonotes} 
\usepackage{xr}
\usepackage{hyperref}
\hypersetup{colorlinks=true, urlcolor=blue,
  linkcolor=blue, citecolor=blue}
\usepackage{subcaption}
\newcommand{\vect}[1]{\mathbf{#1}}

\newcommand*\dif{\mathop{}\!\mathrm{d}}

\begin{document}
\title{Numerical test of the Edwards conjecture shows that all packings are equally probable at jamming}

\author{Stefano Martiniani} 
\email{sm958@cam.ac.uk}
\affiliation{Department of Chemistry, University of Cambridge,
  Lensfield Road, Cambridge, CB2 1EW, UK}
\author{K. Julian Schrenk}
\affiliation{Department of Chemistry, University of Cambridge,
  Lensfield Road, Cambridge, CB2 1EW, UK}
\author{Kabir Ramola} 
\affiliation{Martin Fisher School of Physics,
  Brandeis University, Waltham, MA 02454, USA}
\author{Bulbul Chakraborty} 
\affiliation{Martin Fisher School of
  Physics, Brandeis University, Waltham, MA 02454, USA}
\author{Daan Frenkel}
\affiliation{Department of Chemistry, University of Cambridge,
  Lensfield Road, Cambridge, CB2 1EW, UK}

\maketitle
{\bf In the late 1980s, Sam Edwards proposed a possible
  statistical-mechanical framework to describe the properties of
  disordered granular materials. A key assumption underlying the
  theory was that all jammed packings are equally likely.  In the
  intervening years it has never been possible to test this bold
  hypothesis directly.  Here we present simulations that provide
  direct evidence that at the unjamming point, all packings of soft
  repulsive particles are equally likely, even though generically,
  jammed packings are not. Typically, jammed granular systems are observed precisely at the unjamming point since grains are not very compressible. Our results therefore support Edwards'
  original conjecture.  We also present evidence that at unjamming the
  configurational entropy of the system is maximal.}

In science, most breakthroughs cannot be derived from known physical
laws: they are based on inspired conjectures~\cite{Laughlin}.
Comparison with experiment of the predictions based on such a
hypothesis allows us to eliminate conjectures that are clearly wrong.
However, there is a distinction between testing the consequences of a
conjecture and testing the conjecture itself.  A case in point is
Edwards' theory of granular media. In the late 1980s, Edwards and
Oakeshott~\cite{Edwards} proposed that many of the physical properties
of granular materials (`powders') could be predicted using a
theoretical framework that was based on the assumption that all
distinct packings of such a material are equally likely to be
observed. The logarithm of the number of such packings was postulated
to play the same role as entropy does in Gibbs' statistical-mechanical
description of the thermodynamic properties of equilibrium
systems. However, statistical-mechanical entropy and granular entropy
are very different objects.  Until now, the validity of Edwards'
hypothesis could not be tested directly -- mainly because the number of
packings involved is so large that direct enumeration is utterly
infeasible -- and, as a consequence, the debate about the Edwards
hypothesis has focused on its consequences, rather than on its
assumptions. Here we present results that show that now, at last, it
is possible to test Edwards' hypothesis directly by numerical
simulation. Somewhat to our own surprise, we find that the hypothesis
appears to be correct precisely at the point where a powder is just at
the (un)jamming threshold. However, at higher densities, the
hypothesis fails.  At the unjamming transition, the configurational
entropy of jammed states appears to be at a maximum.

The concept of `ensembles' plays a key role in equilibrium statistical
mechanics, as developed by J. Willard Gibbs, well over a century
ago~\cite{GibbsBook}.  The crucial assumption that Gibbs made in order
to arrive at a tractable theoretical framework to describe the
equilibrium properties of gases, liquid and solids was that, at a
fixed total energy, every state of the system is equally likely to be
observed. The distinction between, say, a liquid at thermal
equilibrium and a granular material is that in a liquid, atoms undergo
thermal motion whereas in a granular medium (in the absence of outside
perturbations) the system is trapped in one of many (very many) local
potential energy minima. Gibbsian statistical mechanics cannot be used
to describe such a system. The great insight of Edwards was to propose
that the collection of all stable packings of a fixed number of
particles in a fixed volume might also play the role of an `ensemble'
and that a statistical-mechanics like formalism would result if one
assumed that all such packings were equally likely to be observed,
once the system had settled into a mechanically stable `jammed' state.
The nature of this ensemble has been the focus of many
studies~\cite{Edwards,Grinev-Edwards, Bi15,Baule16}.

Jamming is ubiquitous and occurs in materials of practical importance,
such as foams, colloids and grains when they solidify in the absence
of thermal fluctuations.  Decompressing such a solid to the point
where it can no longer achieve mechanical equilibrium leads to
unjamming.  Studies of the unjamming transition in systems of
particles interacting via soft, repulsive potentials have shown that
this transition is characterised by power-law scaling of many physical
properties ~\cite{olson_teitel, wyart, silbert_liu_nagel,
  goodrich16,ramola_chakraborty_2016_1, ramola_chakraborty_2016_2}.
However, both the exact nature of the ensemble of jammed states and
the unjamming transition remains unclear.

\begin{figure}
 \includegraphics[width=\columnwidth]{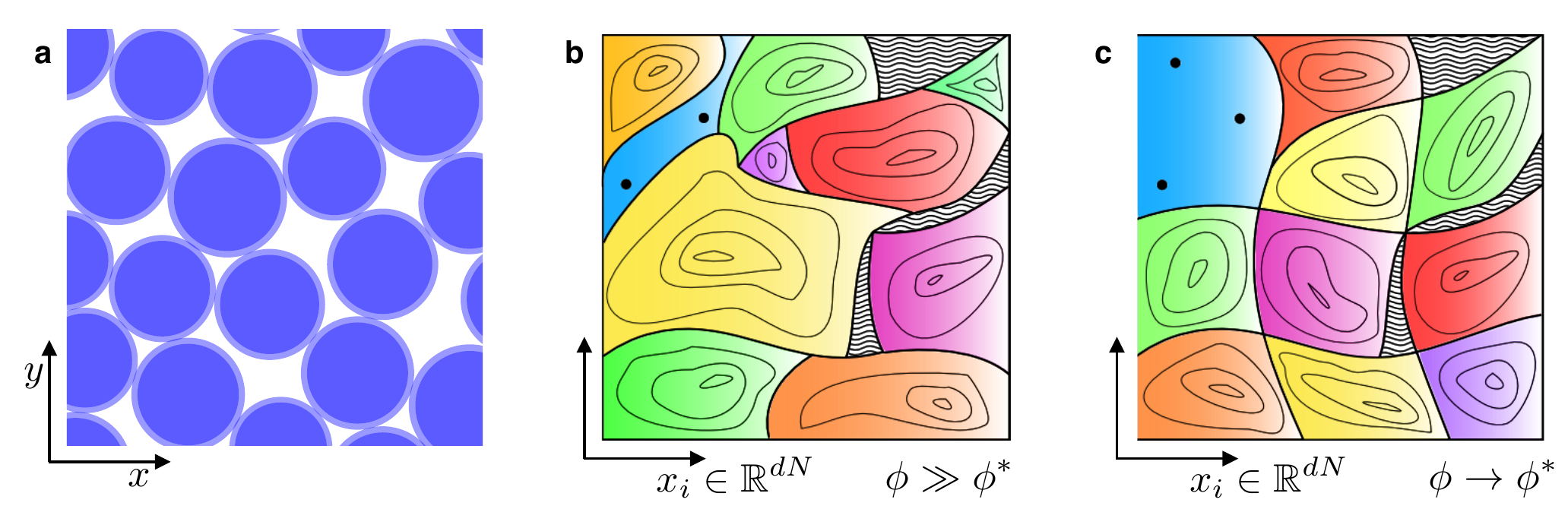}
    \caption{\label{fig::snapshot} Snapshot of the system studied and illustration of the associated energy landscape at different volume fractions. (a) Snapshot of a jammed packing of
      disks with a hard core (dark shaded regions) plus soft repulsive
      corona (light shaded regions). (b)-(c) Illustration of
      configurational space for jammed packings. The dashed regions
      are inaccessible due to hard core overlaps. Single coloured
      regions with contour lines represent the basins of attraction of
      distinct minima. The dark blue region with solid dots indicates
      the coexisting unjammed fluid region and hypothetical marginally
      stable packings, respectively. The volume occupied by the fluid
      $V_{\text{unj}}$ is significant only for finite size systems at
      or near unjamming. When $\phi \gg \phi^*$ (b) the distribution
      of basin volumes is broad but as $\phi \to \phi^*$ (c) the
      distribution of basin volumes approaches a delta function satisfying Edwards' hypothesis.}
\end{figure}

\begin{figure}
 \includegraphics[width=\columnwidth]{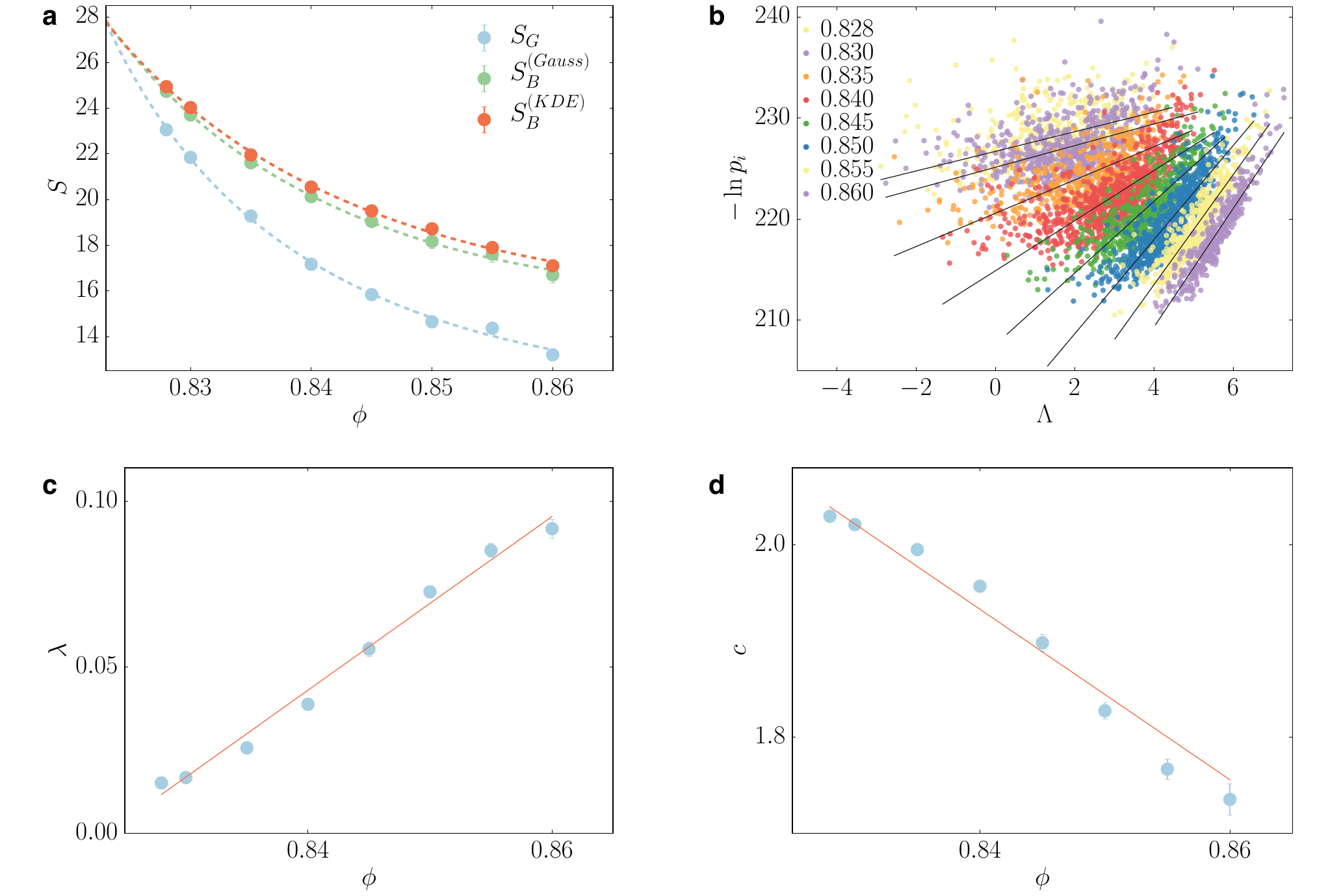}
    \caption{\label{fig::entropy} Numerical results obtained by basin volume calculations for jammed packings of $N=64$ disks with a hard core and a soft shell. (a) Gibbs entropy $S_G$ and
      Boltzmann entropy $S_B$ as a function of volume fraction. $S_B$
      is computed both parametrically by fitting $\mathcal{B}(f)$ with
      a generalised Gaussian function (`Gauss') and non-parametrically
      by computing a Kernel Density Estimate (`KDE') as in
      Ref.~\cite{Martiniani16}. Dashed curves are a second order
      polynomial fit. (b) Scatter plot of the negative log-probability
      of observing a packing, $-\ln p_i = F_i + \ln V_J(\phi)$, where
      $V_J$ is the accessible fraction of phase space (see Methods) as a
      function of log-pressure, $\Lambda$. Black solid lines are lines
      of best fit computed by linear minimum mean square error using a
      robust covariance estimator and bootstrap (see Methods). (c) Slopes $\lambda(\phi)$
      and (d) intercepts $c(\phi)$ of linear fits for
      Eq.~\ref{eq::power-law}. Solid lines are lines of best fit and
      error bars refer to the standard error computed by bootstrap \cite{Efron79}.}
\end{figure}

In this letter, we report a {\it direct test} of the Edwards
conjecture, using a numerical scheme for computing basin volumes of
distinct jammed states (energy minima) of $N$ polydisperse,
frictionless disks held at a constant packing fraction $\phi$.
Uniquely, our numerical scheme allows us to compute $\Omega$, the
number of distinct jammed states, and the individual probabilities
$p_{i \in \{1, \dots, \Omega\}}$ of each observed packing to
occur. Fig.~\ref{fig::snapshot}a shows a snapshot of a section of the
system, consisting of particles with a hard core and a soft shell.  We
obtain jammed packings by equilibrating a hard disk fluid and
inflating the particles instantaneously to obtain the desired
soft-disk volume fractions ($\phi$), followed by energy minimization
(see Methods).  The minimization procedure finds individual stable packings
with a probability $p_i$ proportional to the volume $v_i$ of their
basin of attraction. Averages computed using this procedure,
represented by $\langle \dots \rangle_{\mathcal{B}}$, would then lead
to a bias originating from the different $v_i$'s. Recent advances in
numerical methods~\cite{Xu11, Asenjo14,Martiniani16, Martiniani16b}
have now enabled direct computation of $v_i$, and therefore, an
unbiased characterization of the phase space. A summary of the technique is provided in Methods.
 
We report a detailed analysis of the distribution of $v_i$ for a fixed
number of disks $N = 64$ (all maximum system sizes in our study were
set by the current limits on computing power). We compute $v_i$ using
a thermodynamic integration scheme~\cite{Xu11, Asenjo14,Martiniani16,
  Martiniani16b}, and compute the average basin volume $\langle v
\rangle (\phi)$. The number of jammed states is, explicitly, $\Omega
(\phi) = V_J (\phi)/\langle v \rangle (\phi)$, where $V_J (\phi)$ is
the total available phase space volume at a given $\phi$.  A
convenient way to check equiprobability is to compare the Boltzmann
entropy $S_B = \ln \Omega - \ln {N !}$, which counts all packings with
the same weight, and the Gibbs entropy $S_G = - \sum_i^{\Omega} p_i
\ln p_i - \ln {N !}$ \cite{swendsen2006statistical, frenkel2014why, cates2015celebrating}. The Gibbs entropy satisfies $S_G \leq S_B$, saturating the bound when all $p_i$ are equal: $p_{i \in \{1, \dots,
  \Omega\}} = 1/\Omega$.  As shown in Fig.~\ref{fig::entropy}a, $S_G$
approaches $S_B$ from below as $\phi \to \phi^{*_{(S)}}_{N=64} \approx
0.82$. Fig.~\ref{fig::snapshot}b-c schematically illustrates the
evolution of the basin volumes as the packing fraction is reduced.

To characterize the distribution of basin volumes, we analyse the
statistics of $v_i$ along with the pressure $P_i$ of each packing. It
is convenient to study $F_i \equiv -\ln v_i$ as a function of
$\Lambda_i \equiv \ln P_i$.  As shown in Fig.~\ref{fig::entropy}b, we
observe a strong correlation between $F_i$ and $\Lambda_i$ which we quantify by fitting the data to a bivariate Gaussian distribution. The conditional expectation of F given $\Lambda$ then yields a linear relationship (denoted by solid lines in Fig.~\ref{fig::entropy}b) such that $\langle F \rangle_{\mathcal{B}}
(\phi;\Lambda) \propto \lambda(\phi)\Lambda$, where $\langle F
\rangle_{\mathcal{B}} (\phi;\Lambda)$ represents the average over all
basins at a given $\Lambda$. Previous studies at higher packing fractions~\cite{Martiniani16} indicate that this relationship is preserved in the thermodynamic limit. Defining $f=F/N$, we have (see Methods for
details):
\begin{equation}
\label{eq::power-law}
\begin{aligned} \langle f \rangle_{\mathcal{B}} (\phi; \Lambda) = &
  \lambda(\phi) \Lambda + c(\phi)\\ = & \lambda(\phi) \Delta \Lambda +
  \langle f \rangle_{\mathcal{B}} (\phi) ~,
\end{aligned}
\end{equation}
 where $\Delta \Lambda = \Lambda - \langle \Lambda
 \rangle_{\mathcal{B}} (\phi)$. For Edwards' hypothesis to be valid, we require that in the thermodynamic limit (i) the distribution of volumes approaches a Dirac delta, which follows immediately from the fact that the variance $\sigma^2_f \sim 1/N$ \cite{Asenjo14} (see SI) and (ii) $F_i$ needs to be independent of $\Lambda_i$, as well as of all other structural observables (see SI), and therefore $\lambda(\phi)$ must necessarily vanish.  As can be seen from Fig.~\ref{fig::entropy}c-d, within the range of volume fractions studied, $\lambda(\phi)$ decreases but saturates to a minimum as $\phi \to \phi_{N=64}^{*_{(\lambda)}}$.  We argue below that the saturation is a finite size effect. An extrapolation using the linear regime in Fig.~\ref{fig::entropy}c indicates that $\lambda \rightarrow 0$ at packing fraction $\phi^{*_{(\lambda)}}_{N=64}=0.82 \pm 0.07$, remarkably close to where our extrapolation yields $S_G=S_B$. The analysis of basin volumes, therefore,  strongly suggests that equiprobability is approached only at a characteristic packing fraction and that  the vanishing of $\lambda(\phi)$ can be used to estimate the point of equiprobability.
 
 We next show that $\lambda(\phi)$ does indeed tend to zero in the thermodynamic limit. We use the
 fluctuations $\sigma^2_f$, $\sigma^2_{\Lambda}$, and the covariance
 $\sigma^2_{f \Lambda}$, obtained from the elements of the covariance
 matrix $\hat{\bf \sigma} = ((\sigma_{f}^2,
 \sigma_{f\Lambda}^2),(\sigma_{f\Lambda}^2, \sigma_{\Lambda}^2))$ of
 the joint distribution of $f$ and $\Lambda$ (see Methods for details), to
 define $\lambda$ and $c$ as:

\begin{align}
\label{eq::slope_intercept}
\begin{split}
\lambda(\phi) \equiv
&\frac{\sigma_{f\Lambda}^2(\phi)}{\sigma_{\Lambda}^2(\phi)}, \\ c(\phi)
\equiv & \langle f \rangle_{\mathcal{B}} (\phi) -
\frac{\sigma_{f\Lambda}^2(\phi)}{\sigma_{\Lambda}^2(\phi)} \langle
\Lambda \rangle_{\mathcal{B}} (\phi).
\end{split}
\end{align}

\begin{figure}
\begin{subfigure}{0.5\textwidth}
  \centering
  \includegraphics[width=\linewidth]{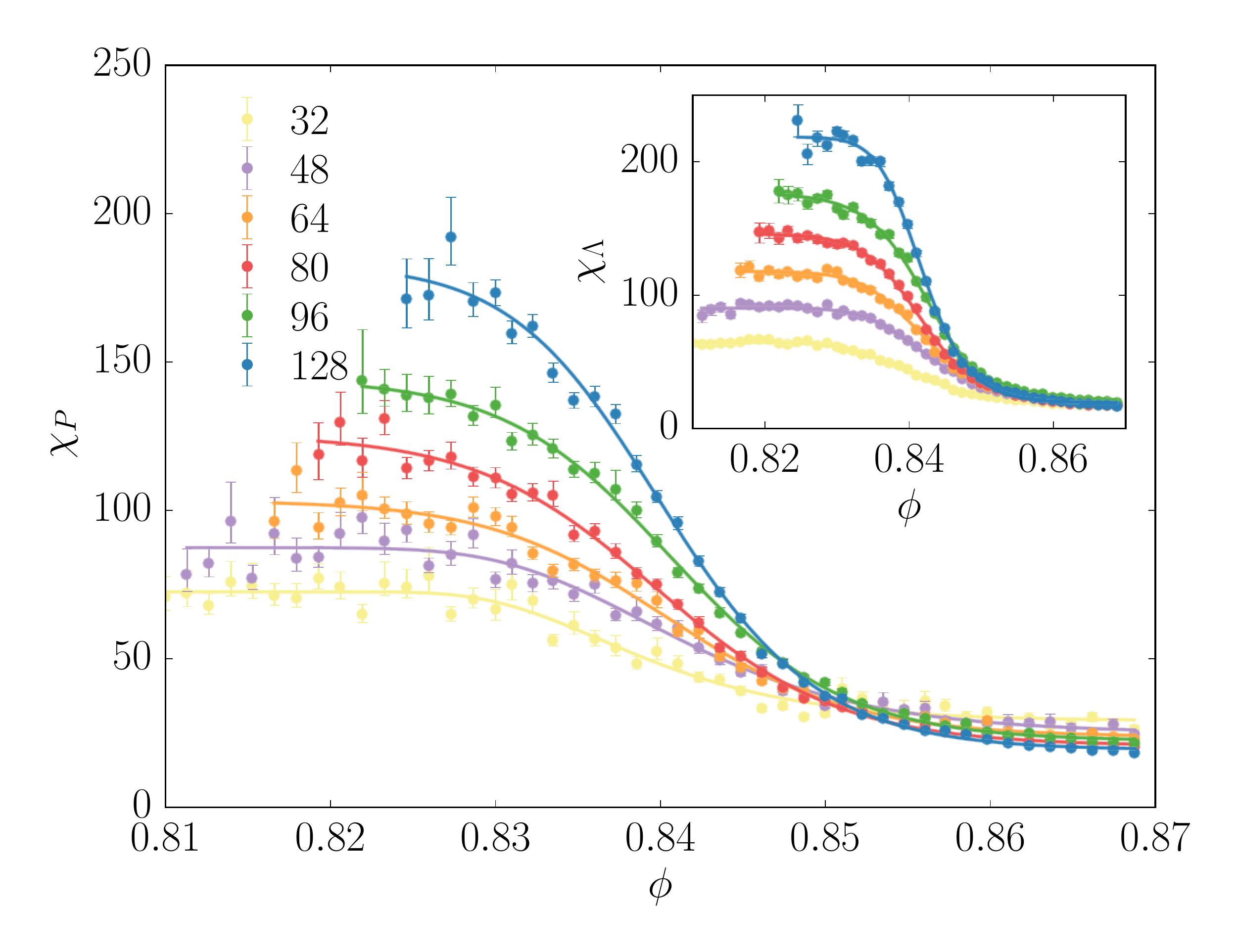}
\end{subfigure}%
\begin{subfigure}{0.5\textwidth}
  \centering
    \includegraphics[width=\linewidth]{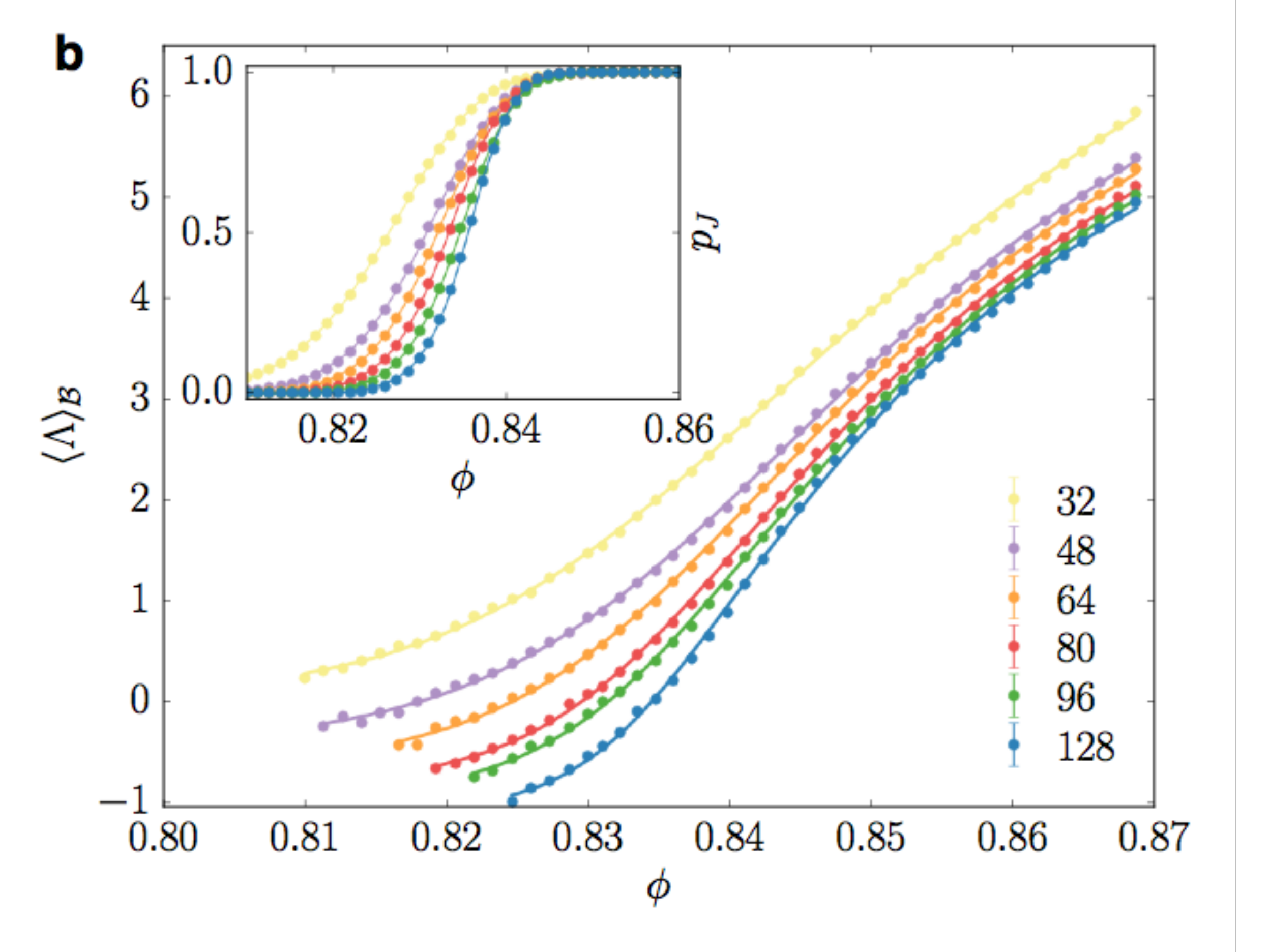}
\end{subfigure}
\caption{\label{fig::lnphi_lnpvar} Finite size scaling analysis. (a) $\chi_P \equiv N
  \sigma^2(P/\langle P \rangle_{\mathcal{B}})$ and (inset)
  $\chi_\Lambda \equiv N \sigma^2_\Lambda$, plotted as a function of
  volume fraction $\phi$. By finite size scaling (see SI) we show that
  the curves diverge in the thermodynamic limit as $\phi \rightarrow
  \phi^{J/*}$, implying $\phi^*_{N\to \infty} = \phi^J_{N \to \infty}
  = 0.841(3)$, see main text for discussion. For $\phi \gg
  \phi^{J}_N$, $\chi_P$ approaches a constant value indicating the
  absence of extensive correlations far from the transition. (b) Observed
  average log-pressure $\langle \Lambda \rangle_\mathcal{B}$ and
  (inset) probability of obtaining a jammed packings by our protocol,
  as a function of volume fraction $\phi$. By finite size scaling (see
  SI) we show that $\langle \Lambda \rangle_\mathcal{B} \to -\infty$
  as $\phi \to \phi_{N\to\infty}^{J_{(\Lambda)}}=0.841(3)$ and $p_J$ collapses for
  $\phi \to \phi_{N\to\infty}^{J_{(p_J)}}=0.844(2)$, thus locating the unjamming
  point. Error bars, computed by BCa bootstrap \cite{Efron87}, refer
  to $1\sigma$ confidence intervals. Solid lines are generalised
  sigmoid fits of the form $f(\phi) = a -
  (a-b)/(1+\exp(-w\Delta\phi))^{1/ u}$. We only show values of $\phi$
  where the probability of finding a jammed packing is at least $1\%$,
  so that the observables are computed over sufficiently large
  sample sizes.}
\end{figure}

From Fig.~\ref{fig::entropy}b we observe that the decrease of $\lambda$ is driven by $\sigma^2_{\Lambda}$ increasing to a maximum, while $\sigma^2_f$ and $\sigma^2_{f \Lambda}$ decrease (see Fig.~S2 of SI). We expect
the main features of these distributions to be preserved as the system
size $N$ is increased \cite{Martiniani16}, which suggests that for
larger $N$, where basin volume calculations are still intractable for multiple densities, the maximum in $\sigma^2_{\Lambda}$ can be used to identify
$\phi^*_N$. We have directly measured $\chi_{\Lambda} = N \sigma^2_\Lambda$
using our sampling scheme -- that samples packings with probability proportional to the volume of their basin of attraction -- for systems of up to $N = 128$ disks (see
inset of Fig.~\ref{fig::lnphi_lnpvar}a) and finite size scaling
indicates that $\chi_{\Lambda}$ diverges as $\phi \to \phi^{*_{(\Lambda)}}_{N \to
  \infty} = 0.841(3)$ (see SI). The saturation of $\lambda$ to a minimum as $\phi \to \phi^*_{N}$, for small $N$, is determined by the fact that $\chi_{\Lambda}$ only diverges in the thermodynamic limit, a detailed discussion is given in Methods.

Interestingly, we find evidence that in the thermodynamic limit, the
point of equiprobability $\phi^*_{N\to \infty}$, coincides with the
point at which the system unjams, $\phi^J_{N \to \infty}$. We use two
characteristics of the unjamming transition to locate $\phi^J_{N \to
  \infty}$ (i) the average pressure of the packings goes to zero, and
therefore $\langle \Lambda \rangle \to - \infty$ (see Fig
\ref{fig::lnphi_lnpvar}b) and (ii) the probability of finding jammed
packings, $p_J$, goes to zero (see inset of Fig
\ref{fig::lnphi_lnpvar}b).  A scaling analysis indicates that
$\langle \Lambda \rangle \to - \infty$ as $\phi^{J_{(\Lambda)}}_{N\to \infty} =
0.841(3)$, and $p_J \to 0$ as $\phi^{J_{(p_J)}}_{N \to \infty}=0.844(2)$ (see
SI).  We thus find that $\phi^*_{N\to \infty} = \phi^J_{N \to \infty}$
within numerical error and up to corrections to finite size scaling
\cite{Vagberg11}.  Our simulations therefore lead to the surprising
conclusion that the Edwards conjecture appears to hold precisely at
the (un)jamming transition. We note that our earlier simulations, which were performed at densities much above jamming \cite{Asenjo14, Martiniani16}, did not support the equiprobability hypothesis. The earlier simulations were in fact far enough away from unjamming that the emergence of equiprobability at this point could not be anticipated. Our earlier findings, therefore, do not contradict the more recent ones and are completely consistent with these.

Why is $\chi_{\Lambda}$ related to the unjamming transition?  As the
particles interact via purely repulsive potentials, the pressure $P$ is strictly
positive, which implies that the fluctuations of $P$ have a floor and
go to zero at unjamming.  The {\it relative} fluctuations $\chi_{P}
\equiv N \sigma^2 \left(P/\langle P \rangle_{\mathcal{B}}\right)$, can
be non-zero, and a diverging $\chi_P$ would then imply a diverging
$\chi_\Lambda$.  Because of the bounded nature of $P$ \cite{henkes_PRE79, henkes_PRE80, Tighe_thesis}, however, $\chi_P$ can {\it only} diverge at the unjamming transition where
$\langle P \rangle_{\mathcal{B}} \to 0$ (see Methods).  We find that
$\chi_P$ does diverge (Fig.~\ref{fig::lnphi_lnpvar}a) and finite size
scaling yields $\phi^{*_{(P)}}_{N \to \infty} = 0.841(3)$, in agreement with what has been found for $\chi_\Lambda$. Returning to the $N = 64$ case that we have analysed using the basin volume
statistics, we find that both $\chi_P$ and $\chi_\Lambda$ saturate to
their maximum values over similar ranges of $\phi$ and our estimate
$\phi^*_{N=64} \approx 0.82$ where $S_G=S_B$ and $\lambda \rightarrow 0$,
falls in this region.  In addition, the average number of contacts
$\langle z \rangle_{\mathcal{B}}(\phi^*_{N=64}) = 4.1 \pm 0.2$ is
close to the isostatic value $z^{(iso)}_{N=64} \equiv 2d - 2/64
\approx 3.97$ \cite{goodrich16} (see SI).

Finally, we note that the states in the generalised Edwards ensemble
\cite{Blumenfeld09, Henkes07, henkes_PRE79, Bi15} characterised by $\phi$
and $P$ have basin volumes that are similar, if not identical, over
the full range of $\phi$ that we have explored (see scatter plot in
Fig.~\ref{fig::entropy}b), indicating that equiprobability in the
stress-volume ensemble \cite{Blumenfeld09, Bi15} is a more robust
formulation of the Edwards hypothesis.  This observation is consistent
with recent experiments \cite{Puckett13}.

Although, the equiprobability of jammed states at a given
packing fraction was posited by Edwards for jammed packings of hard
particles, our analysis shows that for soft particles, the Edwards
hypothesis is valid only for the marginally jammed states at
$\phi^*_{N\to \infty} = \phi^J_{N \to \infty}$, where the jamming
probability vanishes, the entropy is maximised, and relative pressure
fluctuations diverge.  We have shown not only that there exist a
practical `Edwardsian' packing generation protocol, capable of
sampling jammed states equiprobably, but we have uncovered an
unexpected property of the energy landscape for this class of systems.
At this stage, we cannot establish whether the same considerations are
valid in 3D, although the already proven validity of
Eq.~\ref{eq::power-law} in 3D would suggest so \cite{Martiniani16}.
The exact value of the entropy at unjamming, whether finite or not,
also needs to be elucidated.  The implications for `soft' structural
glasses is apparent: at $\phi^J$ the uniform size of the basins
implies that the system, when thermalised, has the same probability of
visiting all of its basins of attraction, hence there are no preferred
inherent structures. This could be a signature of the hard-sphere
transition occurring at the same point \cite{Charbonneau16}.  Our
approach can, therefore, be extended to spin-glasses and related
problems, and it would be clearly very exciting to explore the
analogies and differences between `jamming' in various systems for
which the configuration space can break up into many distinct basins.

\begin{acknowledgments}
S.M. acknowledges financial support by the Gates Cambridge
Scholarship.  K.J.S. acknowledges support by the Swiss National
Science Foundation under Grant No. P2EZP2-152188 and
No. P300P2-161078. D.F. acknowledges support by EPSRC Programme Grant
EP/I001352/1 and EPSRC grant EP/I000844/1. K.R. and B.C. acknowledge
the support of NSF-DMR 1409093 and the W. M. Keck Foundation.
\end{acknowledgments}

S.M, D.F. and B.C. designed the study. S.M. and K.J.S. developed the software and performed the numerical simulations. S.M., K.R., B.C. and D.F. analysed the data and wrote the manuscript.

The data that support the plots within this paper and other findings of this study are available from the corresponding author upon request.

\section{Methods}

\subsection{Packing preparation protocol}
In this section we describe the algorithm that we have used to sample
the phase space of jammed packings. This procedure samples each
configuration proportional to the volume of its basin of attraction.

\subsubsection{Hard sphere fluid sampling}
We start by equilibrating a fluid of N hard disks (that serve as the
cores of the particles with soft outer shells) at a volume fraction
$\phi_{HS}$ in a square box with periodic boundary conditions. The
particle radii are sampled from a truncated Gaussian distribution with
mean $\mu=1$ and standard deviation $\sigma=0.1$.  We achieve
equilibration by performing standard Markov Chain Monte Carlo (MCMC)
simulations consisting of single particle displacements and
particle-particle swaps, as in \cite{Martiniani16}. To assure
statistical independence, we draw fluid configurations every $n_{MC}$
steps, where $n_{MC}$ is (pre-computed by averaging over multiple
simulations) the total number of MCMC steps necessary for each
individual particle to diffuse at least a distance equal to the
largest diameter in the system.

\subsubsection{Soft shells and minimization}
We next take each equilibrated hard disk fluid configuration and
inflate the particles (instantaneously) with a WCA-like soft outer
shell \cite{Weeks71}, to reach the target soft packing fraction
$\phi_\text{SS} > \phi_\text{HS}$. Each hard sphere is inflated
proportionally to its radius, so that the soft sphere radius is given
by
\begin{equation}
    r_{\text{SS}}
    =\left(\frac{\phi_\text{SS}}{\phi_\text{HS}}\right)^{1/d}
    r_{\text{HS}},
\end{equation}
where $d$ is the dimensionality of the box ($2$ in our case), and
$r_{\text{SS}}$ and $r_{\text{HS}}$ are the soft and hard sphere radii
respectively. Clearly, this procedure does not change the
polydispersity of the sample. The radii are identical across volume
fractions and system sizes, and the hard disk fluid density is chosen
so that the radius ratio of hard to soft disks is $(0.88/0.7)^{1/2}
\approx 1.121$.

Next, particle inflation is followed by energy minimization using FIRE
\cite{Bitzek06, Asenjo13}, to produce mechanically stable packings at the desired soft volume fractions $\phi$. This protocol has the advantage of generating packings sampled proportionally to the volume
of their basin of attraction. In our simulations, we considered all
mechanically stable packings, irrespective of the number of
`rattlers'. To guarantee mechanical stability we required that the
total number of contacts is sufficient for the bulk modulus to be
strictly positive, $N_{\text{min}} = d(N_{nr}-1)+1$ \cite{Goodrich12},
where $N_{nr}$ is the number of non-rattlers and $d$ the
dimensionality of the system.

Our implementation of FIRE enforces a maximum step size (set to be equal to the soft shell thickness) and forbids uphill steps by taking one step back every time the energy increases (and restarts the minimizer in the same fashion as the original FIRE implementation). We use a maximum time step $\Delta t_{\text{max}}=1$, although the maximum step size is directly controlled in our implementation. All other parameters are set as in the original implementation \cite{Bitzek06}.

\subsection{HS-WCA potential}

We define the WCA-like potential around a hard core as follows:
consider two spherical particles with a distance between the hard cores
$r_{\text{HS}}$, implying a soft core contact distance
$r_{\text{SS}}=r_{\text{HS}}(1+\theta)$, with $\theta =
(\phi_\text{SS}/\phi_\text{HS})^{1/d}-1$. We can then write a
horizontally shifted hard-sphere plus WCA (HS-WCA) potential as
\begin{equation}
  \label{eq:hswca}
    u_\text{HS-WCA}(r) = 
    \left\{ 
    \begin{array}{l l}
        \infty &\quad r \leq r_{\text{HS}},\\
        \begin{array}{l}
        4\epsilon \left[
          \left(\frac{\displaystyle\sigma(r_{\text{HS}})}{\displaystyle
            r^2-r_{\text{HS}}^2} \right)^{12}
          \right. \\ \left. -\left(\frac{\displaystyle\sigma(r_{\text{HS}})}{\displaystyle
            r^2-r_{\text{HS}}^2} \right)^6 \right] + \epsilon
        \end{array}
        &\quad r_{\text{HS}} < r < r_{\text{SS}}, \\ 0 &\quad r \geq
        r_{\text{SS}}
        \end{array}
    \right.
\end{equation}
where $\sigma(r_{\text{HS}}) = (2\theta +\theta^2)
r_{\text{HS}}^2/2^{1/6}$ guarantees that the potential function and its first derivative go to zero at
$r_{\text{SS}}$. For computational convenience (avoidance of
square-root evaluations), the potential in Eq.~\ref{eq:hswca} differs
from the WCA form in that the inter-particle distance in the
denominator of the WCA potential has been replaced with a difference
of squares. 

A power series expansion of Eq.~\ref{eq:hswca} yields
\begin{equation}
\lim_{r\to r_{\text{SS}}} u_\text{HS-WCA} = \epsilon \left(\frac{12 r_{\text{SS}}}{r_{\text{HS}}^2-r_{\text{SS}}^2}\right)^2 (r-r_{\text{SS}})^2 + O\left((r-r_{\text{SS}})^3\right),
\end{equation}
hence, in the limit of no overlap the pair potential is harmonic.

We numerically evaluate this potential, matching the gradient and
linearly continuing the function $u_\text{HS-WCA}(r)$ for $r \leq
r_{\text{HS}} + \varepsilon$, where $\varepsilon > 0$ is an arbitrary
small constant, such that minimization is still meaningful if hard
core overlaps do occur.

Our choice of potential is based on the fact that (i) the hard cores greatly reduce the amount of configurational space to explore, replacing expensive energy minimizations (to test whether the random walker has stepped outside the basin) with fast hard-core overlap rejections, and (ii) the hard cores exclude high-energy minima (jammed packings) that are not `hard-sphere-like'.
  
\subsection{Total accessible volume}

The basins of attraction of energy minima tile the ``accessible''
phase space (schematically shown in Fig.~1b-c).  This
inaccessible part of the phase space arises due to hard core
constraints and the existence of fluid states (see
e.g.~\cite{Martiniani16}). The total phase space volume is equal to
$V_{\text{box}}^N$. The inaccessible part of this volume arising from
the hard core constraints (shown as hatched areas in Fig.~1) is denoted by $V_{\text{HS}}$, and the part corresponding to the coexisting unjammed fluid states is denoted by $V_{\text{unj}}$ (shown
as blue regions with squares in Fig.~1b-c). $V_{\text{unj}}$
is significant only for finite size systems at or near unjamming.  We
denote the space tiled by the basins of mechanically stable jammed
packings by $V_J$. We then have $V_J = V_{\text{box}}^N -
V_{\text{HS}} - V_{\text{unj}}$. In practice we compute $V_J$ using
the following equation
\begin{equation}
\ln V_J(N, \phi_{\text{HS}}, \phi_{\text{SS}}) = N \ln V_{\text{box}}
- N f_{\text{ex}}(\phi_{\text{HS}}) + \ln p_{J}(\phi_{\text{SS}}),
\end{equation}
where $f_{\text{ex}}(\phi_{\text{HS}})$ is the excess free energy,
\emph{i.e.} the difference in free energy between the hard sphere
fluid and the ideal gas, computed from the Santos-Yuste-Haro (eSYH)
equation of state \cite{Santos99} as in \cite{Martiniani16}, and
$p_{J}(\phi_{\text{SS}})$ is the probability of obtaining a jammed
packing at soft volume fraction $\phi_{\text{SS}}$ with our protocol,
shown in the inset of Fig.~\ref{fig::lnphi_lnpvar}b.

\subsection{Counting by sampling}

We briefly review our approach to computing the number $\Omega$ of distinct jammed packings for a system of $N$ soft disks at volume fraction $\phi$. We prepare packings by the protocol described above, that generates jammed structures (energy minima) with probability $p_i$ proportional to the volume of their basin of attraction $v_i$. We define the probability of sampling the $i$-th packing as
\begin{equation}
p_i = \frac{v_i}{V_J},
\end{equation}
where $V_J$ is the total accessible phase space, such that
\begin{equation}
V_J = \sum_{i=1}^{\Omega} v_i.
\end{equation}
Details of the computation of $v_i$ are discussed in Refs.~\cite{Martiniani16, Martiniani16b}. To find $\Omega$, we make the simple observation
\begin{equation}
\sum_{i=1}^{\Omega} v_i = \frac{\Omega}{\Omega}\sum_{i=1}^{\Omega} v_i = \Omega \langle v \rangle,
\end{equation}
from which it follows immediately that
\begin{equation}
\Omega=\frac{V_J}{\langle v \rangle}.
\end{equation}
The `Boltzmann-like' entropy, suggested in a similar form by Edwards \cite{Edwards}, is then
\begin{equation}
S_B = \ln \Omega - \ln N!
\end{equation}
where the $\ln N!$ correction ensures that two systems in identical macrostates are in equilibrium under exchange of particles \cite{swendsen2006statistical, frenkel2014why, cates2015celebrating}.

Note that $\langle v \rangle$ is the \textit{unbiased} average basin volume (the mean of the unbiased distribution of volumes).  We distinguish between the biased, $\mathcal{B}(\phi; F)$ (as sampled by the packing protocol), and the unbiased, $\mathcal{U}(\phi; F)$, basin log-volumes distributions ($F=-\ln v_{\text{basin}}$).  Since the configurations were sampled proportional to the volume of their
basin of attraction, we can compute the unbiased distribution as
\begin{equation}
    \label{eq::unbiasing}
    \mathcal{U}(\phi; F) = \mathcal{Q}(\phi)
    \mathcal{B}(\phi; F) e^F
\end{equation}
where $\mathcal{Q}(\phi)$ is the normalisation constant, such that
\begin{equation}
    \label{eq::un_bias_integral_c_omega}
    \mathcal{Q}(\phi) = \left [
      \int_{F_\text{min}}^{\infty} \dif F
      \mathcal{B}(\phi; F)e^F \right]^{-1} = \langle v
    \rangle(\phi).
\end{equation}
Since small basins are much more numerous than large ones, and grossly under-sampled, it is not sufficient to perform a weighted average of the sampled basin volumes. Instead, to overcome this problem, one can fit the biased measured basin log-volumes distribution $\mathcal{B}(\phi; F)$ with an analytical (or at least numerically integrable) distribution, and perform the unbiasing via Eq.~\ref{eq::unbiasing} on the best fitting distribution. Different approaches to modelling this distribution give rise to somewhat different analysis methods, which all yield consistent
results as shown in Ref.~\cite{Martiniani16}. In this work we follow Ref.~\cite{Martiniani16} and fit $\mathcal{B}(\phi; F)$ using both a (parametric) generalised Gaussian model \cite{Nadarajah05}, see Eq.~S21 of the SI, and a (non-parametric) kernel density estimate (KDE) with Gaussian kernels \cite{Bishop09, Scikit-learn} and bandwidth selection performed by cross validation \cite{Bowman84, Martiniani16}, yielding consistent results in agreement with Ref.~\cite{Martiniani16}. Before performing the fit we remove outliers from the free energy distribution in an unsupervised manner, as discussed in the ``Data Analysis'' section of the Methods.

No such additional steps are needed to compute the `Gibbs-like' version of the configurational entropy, in fact
\begin{equation}
S_G  = -\sum_{i=1}^\Omega p_i \ln p_i - \ln N! = \sum_{i=1}^\Omega [p_i (-\ln v_i)] + \ln V_J - \ln N! =  \langle F \rangle_{\mathcal{B}} + \ln V_J - \ln N!
\end{equation}
is simply the arithmetic average of the \emph{observed} volumes: The sample mean of $F=-\ln v_{\text{basin}}$ is already correctly weighted because our packing generation protocol generates packings with probability $p_i$.

\subsection{Power-law between pressure and basin volume}
A power-law relationship between the volume of the basin of attraction
of a jammed packing and its pressure was first reported in
\cite{Martiniani16}. In what follows we provide insight into this
expression on the basis of this work's findings. We observe that
distributions of basin negative log-volumes, $F=-\ln
v_{\text{basin}}$, and log-pressures, $\Lambda = \ln P$, are
approximately normally distributed (see Fig.~S1
and S9 of the SI). We therefore expect their joint
probability to be well-approximated by a bivariate Gaussian
distribution $\mathcal{B}(\phi; F,\Lambda) = \mathcal{N}(\vect{\mu},
\hat{\vect{\sigma}})$
\footnote{When listing a function's arguments we place parameters that
  are held constant before the semicolon}, with mean $\vect{\mu} =
(\mu_F, \mu_\Lambda)$ and covariance matrix $\hat{\vect{\sigma}} =
((\sigma_{F}^2, \sigma_{F\Lambda}^2),(\sigma_{F\Lambda}^2,
\sigma_{\Lambda}^2))$ \cite{Bertsekas02}. This is consistent with the
elliptical distribution of points in Fig.~2b. For a
given random variable $X$, with an (observed/biased) marginal
distribution $\mathcal{B}(X)$, the mean is given by $\mu_X(\phi) =
\langle X \rangle_{\mathcal{B}} = \int X \mathcal{B}(\phi; X) \dif X$.
Similarly, the (biased) conditional expectation of $F$ for a given
$\Lambda$ is then \cite{Bertsekas02}
\begin{equation}
\label{eq::fmarginaldist}
\langle F \rangle_{\mathcal{B}(\vect{\Gamma})}(\phi; \Lambda) \equiv \mathbb{E}[ F
  |\phi; \Lambda] = \frac{\sigma_{F\Lambda}^2(\phi)}{\sigma_{\Lambda}^2(\phi)}
(\Lambda - \mu_\Lambda(\phi)) + \mu_F(\phi).
\end{equation}
This is simply the linear minimum mean square error (MMSE) regression
estimator for $F$, \textit{i.e.} the linear estimator $\hat{Y}(X)=aX+b$ that
minimizes $\mathbb{E}[(Y-\hat{Y}(X))^2]$. The expectation of the
dimensionless free energy $\langle F
\rangle_{\mathcal{B}(\vect{\Gamma})} (\phi ; \Lambda) = - \langle \ln
v \rangle_{\mathcal{B}(\vect{\Gamma})} (\phi; \Lambda) \geq - \ln
\langle v \rangle_{\mathcal{B}(\vect{\Gamma})} (\phi; \Lambda) $
\cite{WeissteinJensen} is the average basin negative log-volume at
volume fraction $\phi$ and log-pressure $\Lambda$. Here the average is
also taken over all other relevant, but unknown, order parameters
$\vect{\Gamma}$, such that $\langle F
\rangle_{\mathcal{B}(\vect{\Gamma})}(\phi; \Lambda) = \int \dif
\vect{\Gamma} \mathcal{B}(\vect{\Gamma}) F(\phi;
\vect{\Gamma},\Lambda)$. In other words, we write the expectation of
$F$ at a given pressure as the (biased) average over the unspecified
order parameters $\vect{\Gamma}$. An example of such a parameter would
be some topological variable that makes certain topologies more
probable than others. Note that $F(\phi,\Lambda; \vect{\Gamma})$ is narrowly
distributed around $\mathbb{E}[ F |\phi; \Lambda]$. To simplify the
notation we write $\langle F \rangle_{\mathcal{B}(\vect{\Gamma})}(\phi; \Lambda) \equiv \langle F \rangle_{\mathcal{B}}(\phi; \Lambda)$.  We can thus rewrite the
power-law reported in \cite{Martiniani16} as
\begin{equation}
\label{eq::power-law}
\begin{aligned} 
\langle f \rangle_{\mathcal{B}}(\phi; \Lambda) = &
  \lambda(\phi) \Lambda + c(\phi)\\ = &
  \frac{\sigma_{f\Lambda}^2(\phi)}{\sigma_{\Lambda}^2(\phi)} \Lambda -
  \frac{\sigma_{f\Lambda}^2(\phi)}{\sigma_{\Lambda}^2(\phi)} \mu_\Lambda(\phi) +
  \mu_f(\phi)\\ = & \frac{\sigma_{f\Lambda}^2(\phi)}{\sigma_{\Lambda}^2(\phi)}
  \Delta \Lambda + \mu_f(\phi) 
\end{aligned}
\end{equation}
 where $f = F/N$ is the basin negative log-volume per particle and
 $\lambda\equiv 1/\kappa$ is the slope of the power-law relation,
 which depends crucially on the packing fraction $\phi$. The last
 equality in Eq. \ref{eq::power-law} highlights how $\lambda(\phi)$
 controls the contributions of the fluctuations of the log-pressures
 $\Delta\Lambda \equiv \Lambda - \mu_\Lambda(\phi)$ to changes in the basin negative log-volume. Note that one can rewrite the ratio of fluctuations
 as $\sigma^2_{f\Lambda}/\sigma^2_{\Lambda} = \rho_{f\Lambda} \sigma_f
 / \sigma_\Lambda$ where $\rho_{f\Lambda}=
 \sigma^2_{f\Lambda}/(\sigma_f \sigma_\Lambda)$ is the linear
 correlation coefficient of $f$ and $\Lambda$. Finally, we can gain
 further insight into the power-law dependence by noting that
\begin{align}
\lambda(\phi) \equiv
&\frac{\sigma_{f\Lambda}^2(\phi)}{\sigma_{\Lambda}^2(\phi)} \label{17} \\ 
c(\phi) \equiv & \mu_f(\phi) - \frac{\sigma_{f\Lambda}^2(\phi)}{\sigma_{\Lambda}^2(\phi)}
\mu_\Lambda(\phi) \label{eq::intercept}
\end{align}

\subsection{Data analysis}

\subsubsection{Reduced units}

While presenting data from our computations, we express pressure and
volume in reduced units as $P/P^{\ast}$ and $v/v^{\ast}$
respectively. The unit of volume is given by $v^{\ast} \equiv \pi
\langle r_{HS} ^2\rangle$, where $\langle r_{HS}^2 \rangle$ is the
mean squared hard sphere radius. The unit of pressure is then
$P^{\ast}\equiv \epsilon/v^{\ast}$, where $\epsilon$ is the stiffness
of the soft-sphere potential, defined in Eq. \ref{eq:hswca}. The
pressure is computed as $P = \text{Tr}(\hat{\bf \Sigma})/2V_{\text{box}}$ where
$\hat{\bf \Sigma}$ is the Virial stress tensor and $V_{\text{box}}$ the volume
of the enclosing box.

\subsubsection{Summary of calculations}

For the basin volume calculations we consider systems of $N=64$ disks
sampled at a range of $8$ volume fractions $0.828 \leq \phi \leq 0.86$
and for each $\phi$ we measure the basin volume for about $365<M<770$
samples.

For the finite size scaling analysis of the relative pressure
fluctuations we study system sizes $N=32,48,64,80,96,128$ for $48$
volume fractions in the range $0.81 \leq \phi \leq 0.87$. For each
system size we generate up to $10^5$ hard disk fluid configurations
and compute the pressure for between approximately $10^3$ and $10^4$ jammed
packings (depending on the probability of obtaining a jammed packing
at each volume fraction).

Simulations were performed using the open source libraries PELE
\cite{pele} and MCPELE \cite{mcpele}.

\subsubsection{Outlier detection and robust covariance estimation}

Before manipulating the raw data we remove outliers from the joint
distribution $\mathcal{B}(f,\Lambda)$ following the distance-based
outlier removal method introduced by Knorr and Ng \cite{Knorr98}. This
is applied in turn to each dimension, such that we choose to keep only
those points for which at least $R=0.5$ of the remaining data set is
within $D=4\sigma$ (compared to the much stricter $R=0.9988$,
$D=0.13\sigma$ required to exclude any points further than
$|\mu-3\sigma|$ for normally distributed data \cite{Knorr98}). On our
datasets we find that this procedure removes typically none and at
most $0.8\%$ of all data points.

Mean and covariance estimates of $\mathcal{B}(f,\Lambda)$ are computed
using a robust covariance estimator, namely the Minimum Covariance
Determinant (MCD) estimator \cite{Rousseeuw99, Scikit-learn} with
support fraction $h/n_{\text{samples}}=0.99$. The MCD estimator
defines $\mu_{\text{MCD}}$, the mean of the $h$ observations for which
the determinant of the covariance matrix is minimal, and
$\hat{\vect{\sigma}}_{\text{MCD}}$, the corresponding covariance matrix
\cite{Hubert10}. We use these robust estimates of the location and of
the covariance matrix (computed over $1000$ bootstrap samples \cite{Efron79}) to fit our observations by linear MMSE \cite{Bertsekas02}, see Fig.~2.

Before fitting $\mathcal{B}(f)$ (required to compute $\Omega$), we
perform an additional step of outlier detection based on an elliptic
(Gaussian) envelope criterion constructed using the MCD estimator. We
assume a support fraction $h/n_{\text{samples}}=0.99$ and a
contamination equal to $10\%$ \cite{Scikit-learn}. We compute $S_G$
and $S_B$ from the resulting datasets. The procedure is strictly
unsupervised and allows us to achieve robust fits despite the small
sample sizes. We fit $\mathcal{B}(f)$ using both a (parametric)
generalised Gaussian model \cite{Nadarajah05} and a (non-parametric)
kernel density estimate (KDE) with Gaussian kernels \cite{Bishop09,
  Scikit-learn} and bandwidth selection performed by cross validation
\cite{Bowman84, Martiniani16}.

\subsubsection{Error analysis}
Errors were computed analytically where possible and propagated using
the `uncertainties' Python package
\cite{uncertainties}. Alternatively, intervals of confidence were
computed by bootstrap for the covariance estimation \cite{Efron79} and by BCa bootstrap otherwise using the `scikit-bootstrap' Python package \cite{Efron87, scikit-bootstrap}.

\end{document}


\title{Supplementary Information: Numerical test of the Edwards conjecture shows that all packings become equally probable at jamming}

\author{Stefano Martiniani} 
\email{sm958@cam.ac.uk}
\affiliation{Department of Chemistry, University of Cambridge,
  Lensfield Road, Cambridge, CB2 1EW, UK}
\author{K. Julian Schrenk}
\affiliation{Department of Chemistry, University of Cambridge,
  Lensfield Road, Cambridge, CB2 1EW, UK}
\author{Kabir Ramola} 
\affiliation{Martin Fisher School of Physics,
  Brandeis University, Waltham, MA 02454, USA}
\author{Bulbul Chakraborty} 
\affiliation{Martin Fisher School of
  Physics, Brandeis University, Waltham, MA 02454, USA}
\author{Daan Frenkel}
\affiliation{Department of Chemistry, University of Cambridge,
  Lensfield Road, Cambridge, CB2 1EW, UK}

\maketitle
\beginsupplement

\section{Variance of relative pressures}
In this section we relate the statistics of the log-pressures of the
packings to the {\it relative} pressures. For a given $N$ and $\phi$,
with the set of pressures $\{ P_i \}$, the log-pressures are given by
$\Lambda_i \equiv\ln P_i$ and the relative pressures are $ P_i/\langle P
\rangle$.  The two quantities are then simply related as
\begin{equation}
\ln \left( \frac{P_i}{\langle P \rangle} \right) = \ln P_i - \ln \langle P \rangle = \Lambda_i - \ln \langle P \rangle .
\label{relative_log_relation}
\end{equation}
Using Jensen's inequality \cite{WeissteinJensen}, we have the following bound for the first moment of the log-pressures
\begin{equation}
\langle \Lambda \rangle = \langle \ln P \rangle \le \ln \langle P \rangle.
\label{Jensen_inequality}
\end{equation}
Therefore $\langle P \rangle \to 0$ implies $\langle \Lambda \rangle
\to -\infty$.  In order to relate the means and the variances of $P_i$
and $\Lambda_i \equiv \ln P_i$, we perform the Taylor expansion
\begin{equation}
\ln P = \ln \langle P \rangle + \left. \frac{d \ln P}{d P} \right
\rvert_{P = \langle P \rangle} (P - \langle P \rangle ) + ...
\end{equation}
We next compute the moments to leading order
\begin{eqnarray}
\nonumber \langle \ln P \rangle &\approx& \ln \langle P
\rangle,\\ \sigma^2(\ln P) &\approx& \frac{\sigma^2(P)}{\langle P
  \rangle^2} = \sigma^2 \left( \frac{P}{\langle P \rangle} \right).
\end{eqnarray}
Thus, to first order, the variance of the log-pressures is equal to the variance of the relative pressures.

\subsection{Bounds}
For a fixed $N$ and $\phi$, the pressures of the individual packings are bounded as 
\begin{equation}
0< P_{\rm min} \le P_i \le P_{\rm max}.
\end{equation}
where $P_{\rm max}$ and $P_{\rm min}$ are determined by the packing
fraction $\phi$, independent of system size, and are physically set
limits \cite{henkes_PRE79, henkes_PRE80, Tighe_thesis}.  We can therefore use Popoviciu's inequality on variances, yielding
\begin{equation}
\sigma^2(P) \le \frac{1}{4} \left( P_{\rm max} - P_{\rm min}\right)^2,
\end{equation}
which is also bounded. The relative pressure fluctuations are therefore bounded as
\begin{equation}
\sigma^2\left(\frac{P}{\langle P \rangle}\right) \le \frac{1}{4}
\frac{\left( P_{\rm max} - P_{\rm min}\right)^2}{\langle P \rangle^2}.
\end{equation}
We thus find that the variance of the relative pressures
$\sigma^2\left(\frac{P}{\langle P \rangle}\right)$ can diverge only
when $\langle P \rangle \to 0$, which is precisely where the unjamming
transition occurs.

\subsection{Scaling at $\phi \gg \phi^*$}

First we show that away from a critical point, the relative pressure fluctuations scale as $1/L^2$, where $L=\sqrt{N}$ and $N$ is the number of particles. The internal Virial is defined by $P = \sum_{i=1,N} p_i$, where $p_i$ is the particle level ``pressure'' given by $p_i = \sum_j \sum_{\alpha=1,2} \vect{r}_{i,j}^\alpha \vect{f}_{i,j}^\alpha$ where $\vect{r}_{i,j}^\alpha$ and $\vect{f}_{i,j}^\alpha$ are the contact vectors and contact forces, respectively. The variance of $P/\langle P \rangle$ is
\begin{equation}
\sigma^2(P/\langle P \rangle) = \frac{\langle P^2 \rangle-\langle P \rangle^2}{\langle P \rangle^2} = \frac{\sum_{i=1}^N \sigma^2(p_i) + \sum_{i \neq j} \text{cov}(p_i,p_j)}{\left (\sum_{i=1}^N \langle p_i \rangle \right )^2}.
\end{equation}
When away from a critical point we expect $\sum_{i \neq j} \text{cov}(p_i,p_j)$ to scale subextensively, and the variance of relative pressure fluctuations to be
\begin{equation}
\sigma^2(P/\langle P \rangle) = \frac{\sum_{i=1}^N \sigma^2(p_i)}{\left (\sum_{i=1}^N \langle p_i \rangle \right )^2} \sim \frac{1}{N}
\end{equation}
hence the relative pressure fluctuations away from the critical point will scale as $1/N = 1/L^2$, and 
\begin{equation}
\sigma^2(\Lambda) \approx \sigma^2(P/\langle P \rangle) \sim 1/L^2,
\end{equation}
as can be verified in Fig.~3a of the main text. 

Second, we analyse the covariance of the basin negative log-volume per particle ($F/N=-\ln( v_{\text{basin}})/N$) and the relative pressure fluctuations, $\text{cov}(F/N,P/\langle P \rangle)$. Similarly to the internal Virial we define the particle level basin negative log-volume as $F = \sum_{i=1,N} f_i$. Then we have
\begin{equation}
\text{cov}(F/N,P/\langle P \rangle) = \frac{\langle FP \rangle - \langle F \rangle\langle P \rangle}{N \langle P \rangle} = \frac{\sum_{i=1}^N \text{cov}(f_i,p_i) + \sum_{i \neq j} \text{cov}(f_i,p_j)}{N \sum_{i=1}^N \langle p_i \rangle}.
\end{equation}
From the power-law relation between $F$ and $\Lambda$ we know that away from the critical point $\text{cov}(f_i,p_i) > 0$  and we expect $\sum_{i \neq j} \text{cov}(f_i,p_j)$ to scale subextensively in this region, hence
\begin{equation}
\text{cov}(F/N,P/\langle P \rangle) = \frac{\sum_{i=1}^N \text{cov}(f_i,p_i)}{N \sum_{i=1}^N \langle p_i \rangle} \sim \frac{1}{N},
\end{equation}
therefore
\begin{equation}
\label{eq::cov_rand}
\text{cov}(f, \Lambda) \approx \text{cov}(f, P/\langle P \rangle) \sim 1/L^2, 
\end{equation}
where $f = F/N$. We can thus conclude that the slope of the power-law relation is
\begin{equation}
\lambda_{\phi \gg \phi^*} = \frac{\sigma^2(f,\Lambda)}{\sigma^2(\Lambda)} \sim \mathcal{O}(1),
\end{equation}
In other words, $\lambda$ is independent of system size, a fact that has been verified numerically in Ref.~\cite{Martiniani16}.

\subsection{Scaling as $\phi \to \phi^*$}
Near the critical point, as $\phi \to \phi^*$, the variance of $\Lambda$ follows the scaling form
\begin{equation}
\label{eq::var_lambda_scaling}
\sigma^2(\Lambda) \sim L^{\gamma/\nu-2},
\end{equation}
with $\gamma/\nu \approx 1$ as found by finite size scaling, shown in Fig.~\ref{fig::var_lnp}. While we do not have a finite size scaling collapse for the covariance $\sigma^2(f, \Lambda)$, due to the high computational cost of performing the basin volume calculations for multiple system sizes, we do observe that for $N=64$ the covariance decreases with respect to the ``background'' $1/L^2$ fluctuations as $\phi \to \phi^*$, see Fig.~\ref{fig::biased_pdf_bv_moments}e. Since we expect this to be generic, we do not expect $L^2\sigma^2(f, \Lambda)$ to diverge but rather that
\begin{equation}
\sigma^2(f,\Lambda) \lesssim L^{-2},
\end{equation}
Hence in the limit $\phi \to \phi^*$ we expect that the slope of the power-law relation will be
\begin{equation}
\lambda_{\phi \to \phi^*} = \frac{\sigma^2(f,\Lambda)}{\sigma^2(\Lambda)} = 0.
\end{equation}

\subsection{Relation between scaling exponents}
Starting from Eq.~1 of the main text, we use the fact that $\sigma^2(aX \pm bY) = a^2\sigma^2(X) + b^2\sigma^2(Y) \pm 2 ab~\mathrm{cov}(X,Y)$ to compute the variance of $f = F/N$ to find
\begin{equation}
\sigma^2_f = \lambda^2 \sigma^2_\Lambda = (\sigma^2_{f\Lambda})^2/\sigma^2_\Lambda
\end{equation}
By rearranging this expressions we find that
\begin{equation}
(\sigma^2_{f\Lambda})^2/\sigma^2_f = \sigma^2_\Lambda \sim \left\{
                \begin{array}{ll}
                  L^{-2} \text{ for } \phi \gg \phi^*\\
                  L^{-\zeta} \text{ for } \phi \to \phi^*
                \end{array}
              \right.
\end{equation}
where we have defined $\zeta \equiv 2 - \gamma/\nu \approx 1$ as in Eq.~(\ref{eq::var_lambda_scaling}) and as found by finite size scaling, shown in Fig.~\ref{fig::var_lnp}. For $\phi \to \phi^*$, by assuming scalings $\sigma^2_{f\Lambda} \sim L^{-\eta}$ and $\sigma^2_{f} \sim L^{-\vartheta}$, we find the following relation between scaling exponents
\begin{equation}
2\eta - \vartheta = \zeta
\end{equation}

\begin{figure}[b]
\begin{subfigure}{0.5\textwidth}
  \centering
  \includegraphics[width=\linewidth]{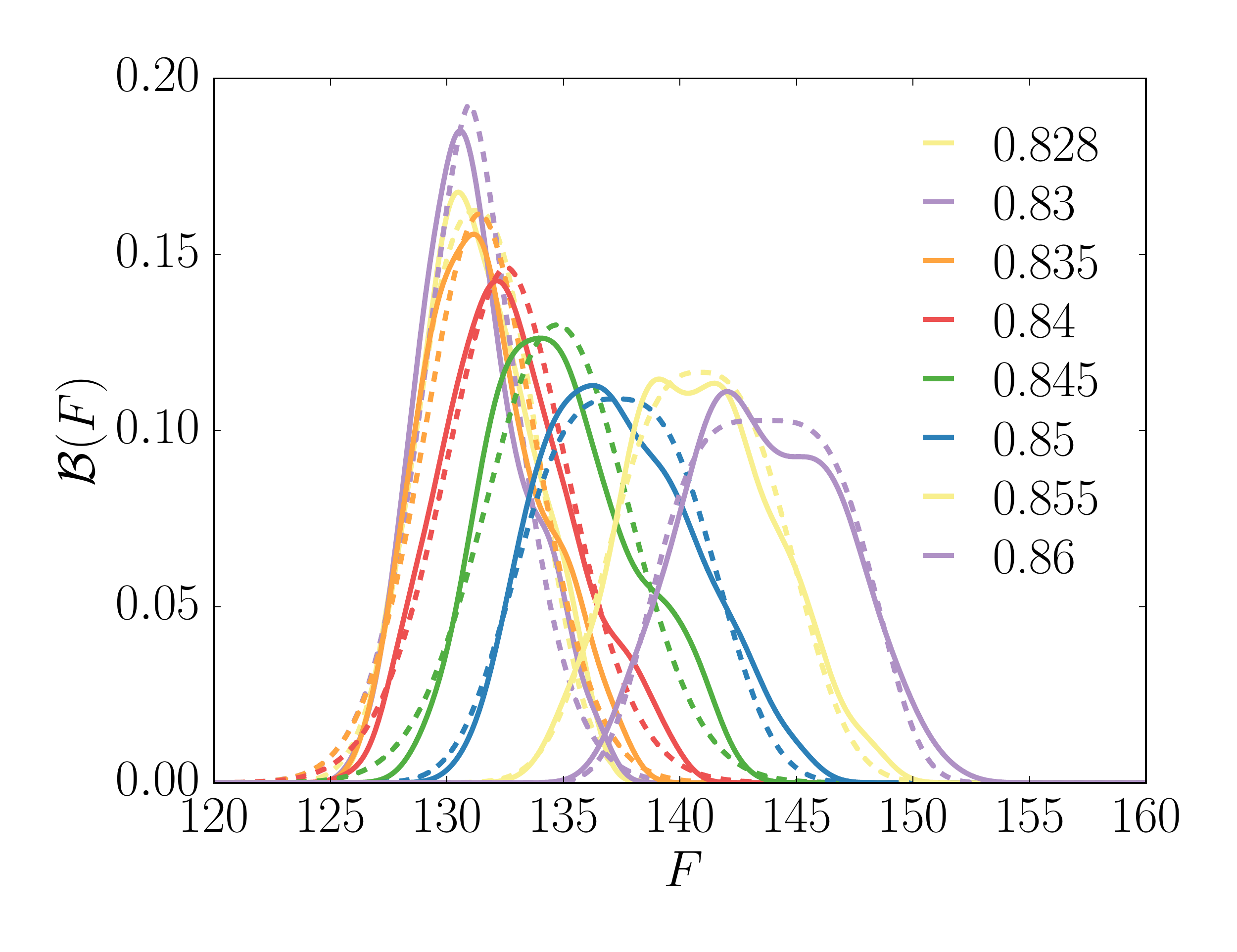}
	\caption{}
\end{subfigure}%
\begin{subfigure}{0.5\textwidth}
  \centering
    \includegraphics[width=\linewidth]{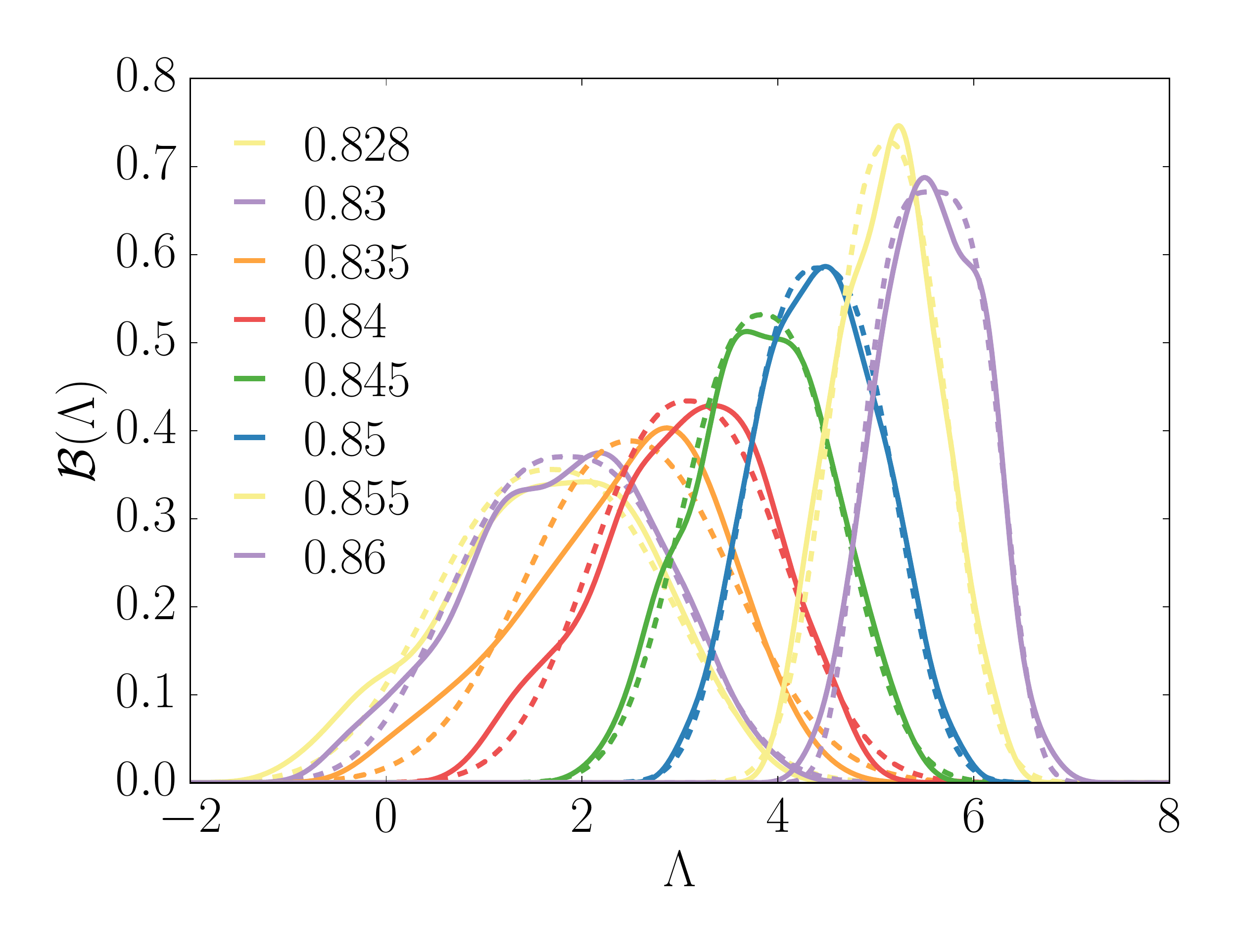}
	\caption{}
\end{subfigure}
\caption{\label{fig::biased_pdf_bv} Observed distribution of the basin
  log-volume $F$ (a) and log-pressure $\Lambda$ (b) for jammed
  packings of $N=64$ HS-WCA polydisperse disks at various volume
  fractions $0.828 \leq \phi \leq 0.86$. Solid lines are Kernel
  Density Estimates and dashed lines are generalised Gaussian fits.}
\end{figure}

\section{Distribution of basin log-volumes}

The distributions of basin negative log-volumes, shown in Fig.~\ref{fig::biased_pdf_bv}, are well represented by a three-parameter generalised Gaussian distribution

\begin{equation}
    \label{eq::gen_gaussian}
    \mathcal{B}(F|\mu_F,\sigma_F,\zeta) \equiv
    \frac{\zeta}{2\sigma\Gamma(1/\zeta)}\exp\left[-\left(\frac{|F-\mu_F|}{\sigma_F}\right)^{\zeta}\right],
\end{equation}
where $\Gamma(x)$ is the gamma function, $\sigma_F$ is the scale parameter, $\zeta$ is the shape parameter and $\mu_F$ is the mean log-volume with variance $\sigma^2 \Gamma(3/\zeta)/\Gamma(1/\zeta)$. In Ref.~\cite{Asenjo13} it is shown that in the limit $N \to \infty$ the shape parameter approaches that of a standard Gaussian distribution, $\zeta = 2$. Since $\sigma_F^2 \sim N$ and $\mu_F \sim N$, we have that
\begin{equation}
e^{-(F-\mu_F)^2/(2\sigma^2)} \sim e^{-N(f-\mu_f)^2} \to \delta(\mu_f) \text{ as } N \to \infty,
\end{equation}
where $\delta$ is the Dirac delta function and $f = F/N$. The distribution of basin volumes thus becomes infinitely narrow in the thermodynamic limit. However, this is not sufficient for the Edwards conjecture to be correct, in fact we also require that the basin volumes are uncorrelated with respect to any structural observables in this limit. In this manuscript we argue that this occurs only as $\phi \to \phi^*$.

\section{Distributions of F and $\Lambda$}
\begin{figure}
\begin{subfigure}{0.5\textwidth}
  \centering
  \includegraphics[width=\linewidth]{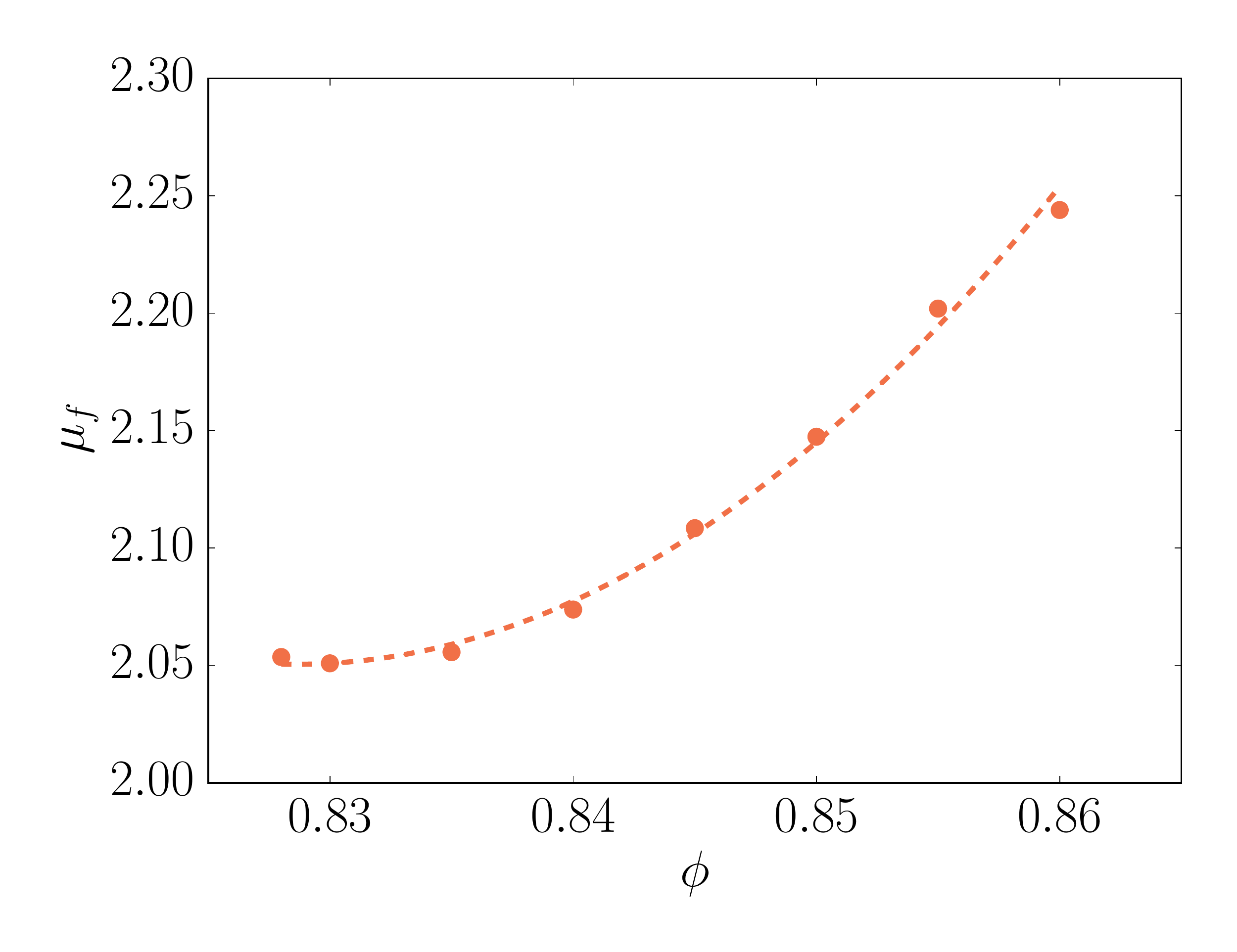}
	\caption{}
\end{subfigure}%
\begin{subfigure}{0.5\textwidth}
  \centering
  \includegraphics[width=\linewidth]{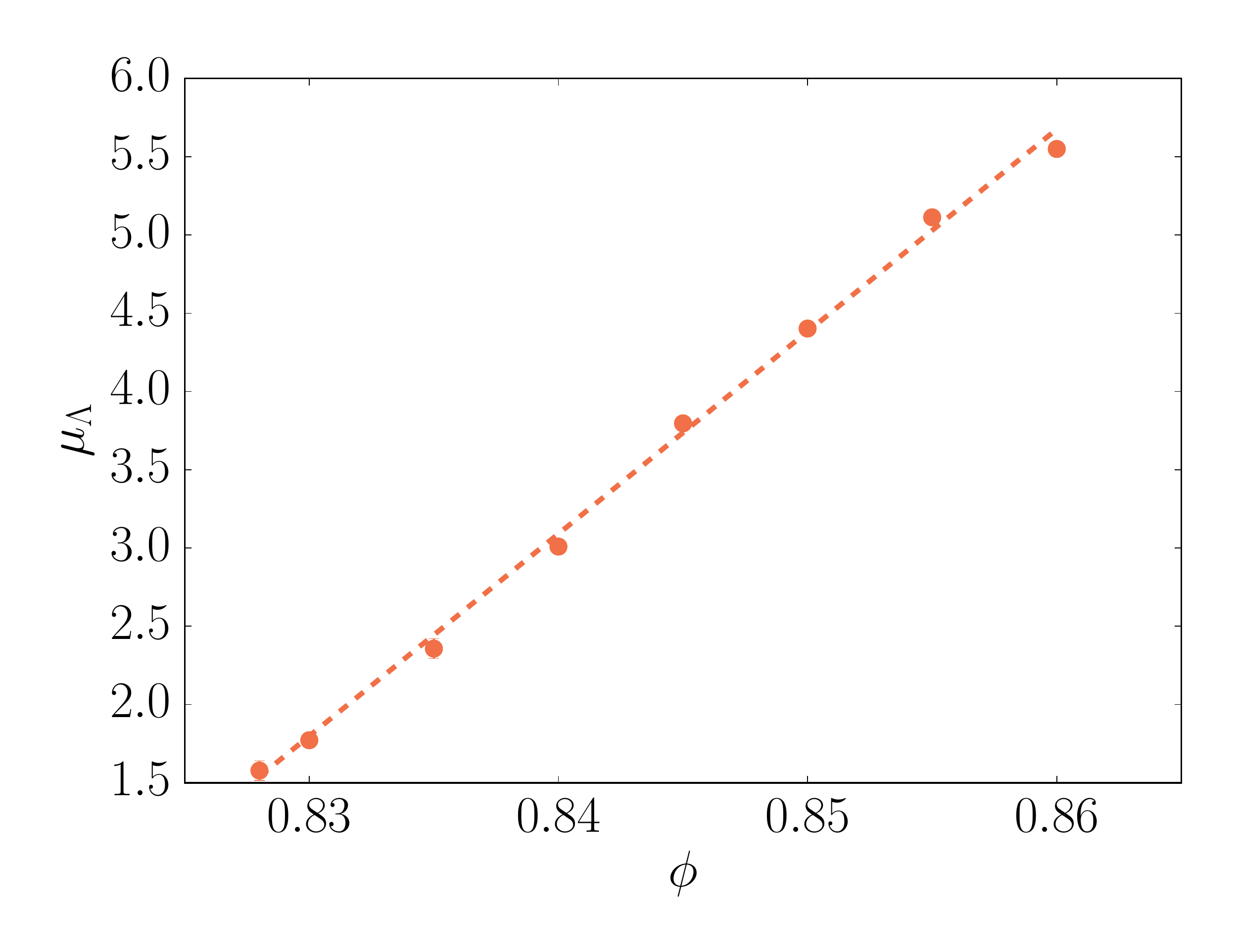}
	\caption{}
\end{subfigure}
\begin{subfigure}{0.5\textwidth}
  \centering
    \includegraphics[width=\linewidth]{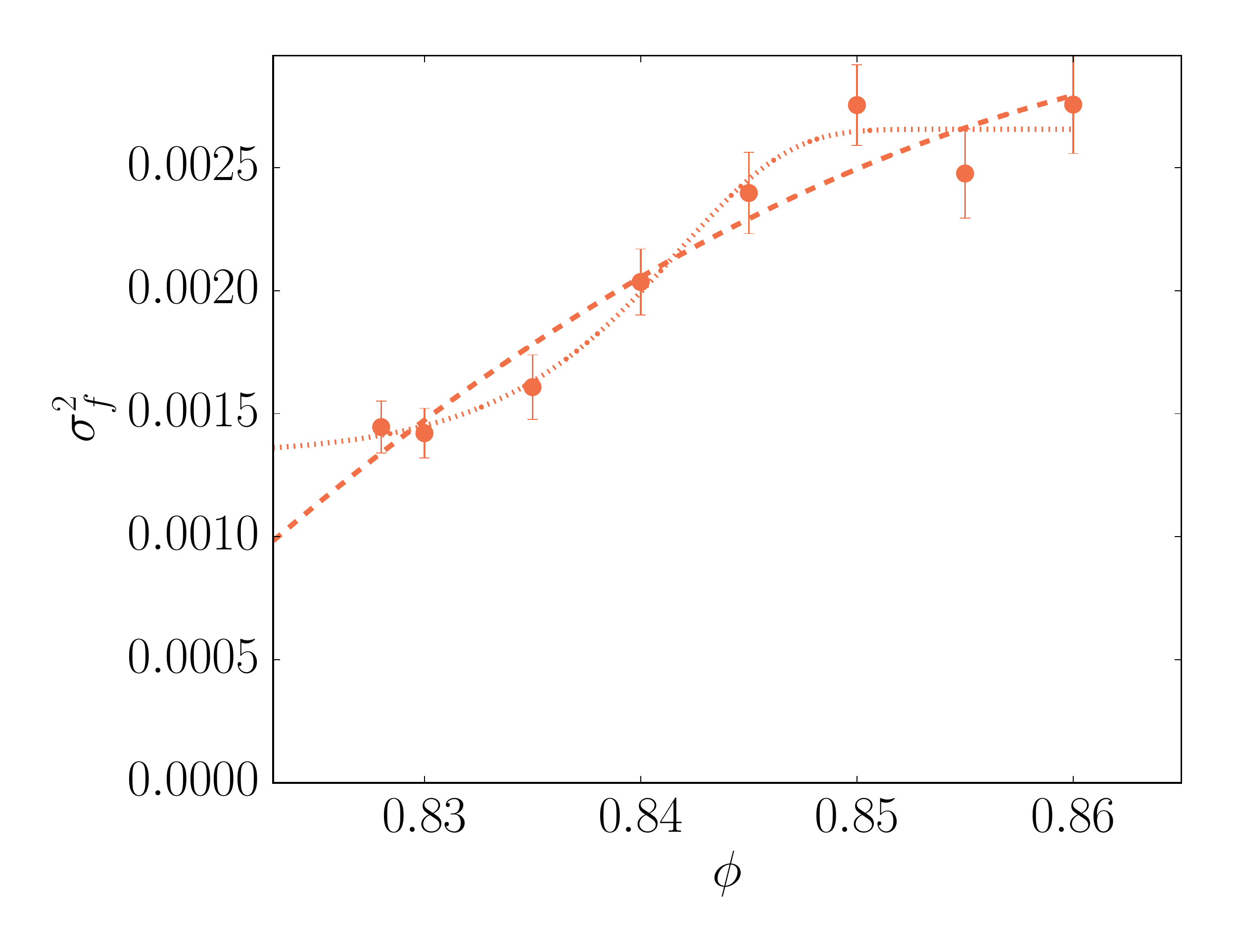}
	\caption{}
\end{subfigure}%
\begin{subfigure}{0.5\textwidth}
  \centering
    \includegraphics[width=\linewidth]{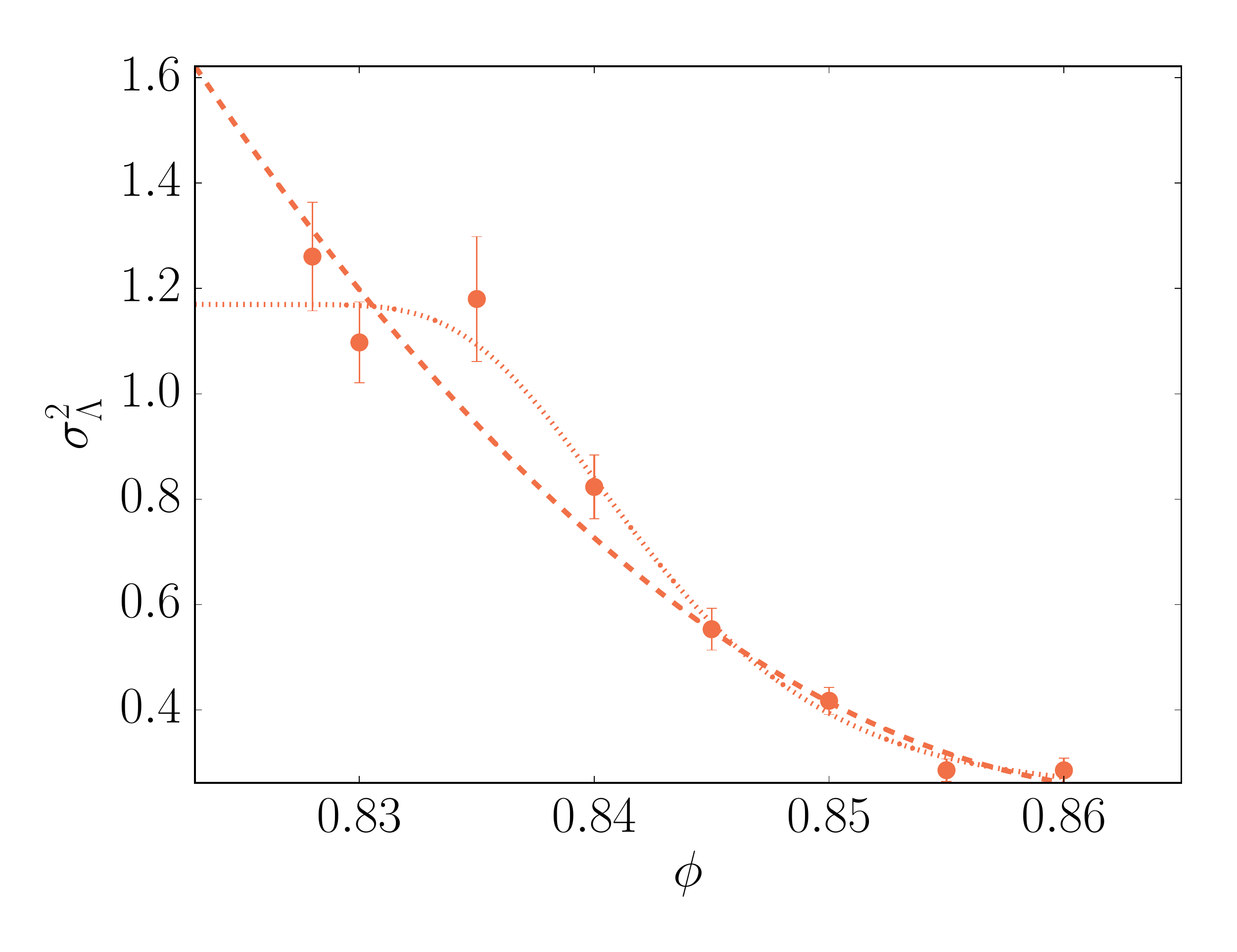}
	\caption{}
\end{subfigure}
\begin{subfigure}{0.5\textwidth}
  \centering
    \includegraphics[width=\linewidth]{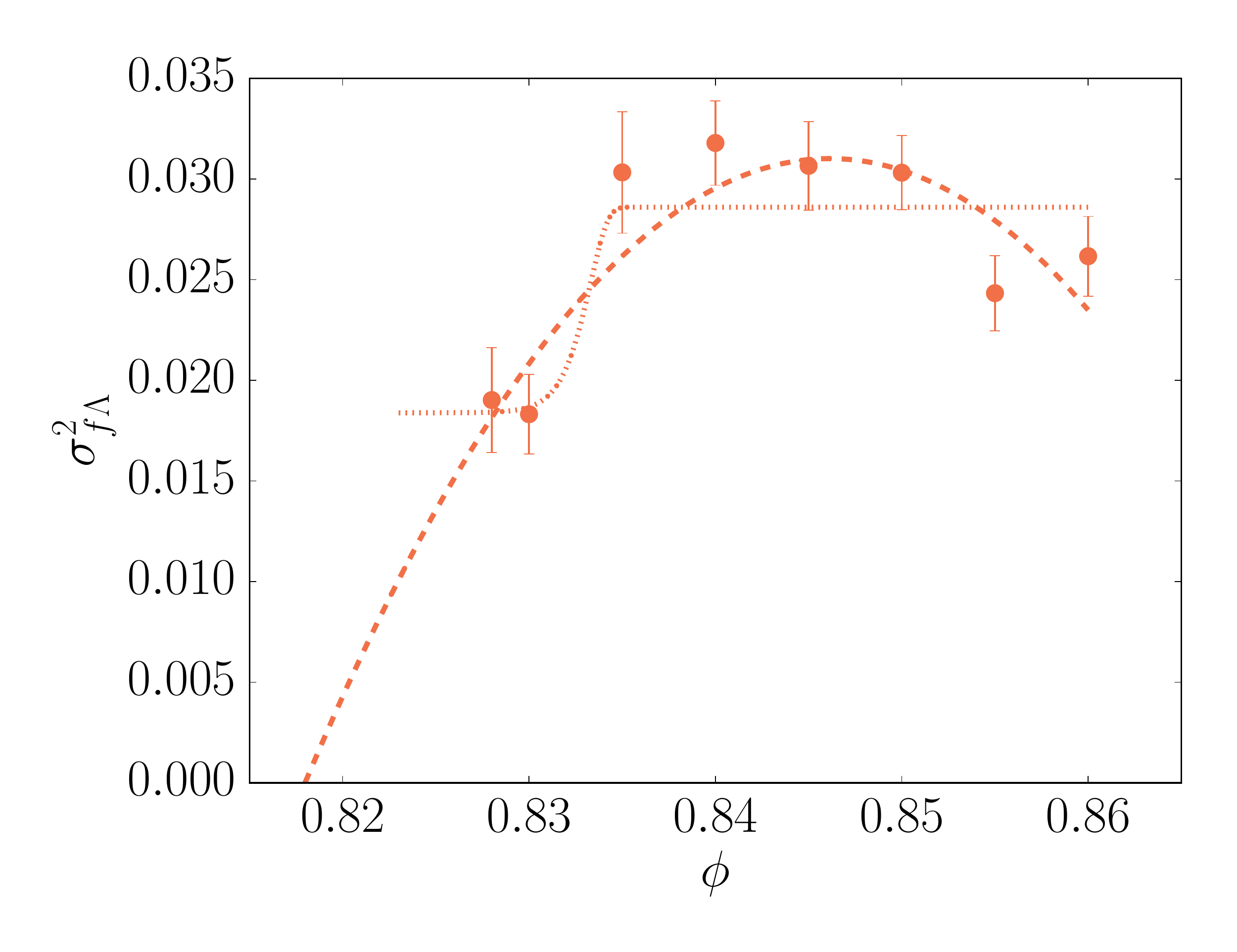}
	\caption{}
\end{subfigure}%
\begin{subfigure}{0.5\textwidth}
  \centering
    \includegraphics[width=\linewidth]{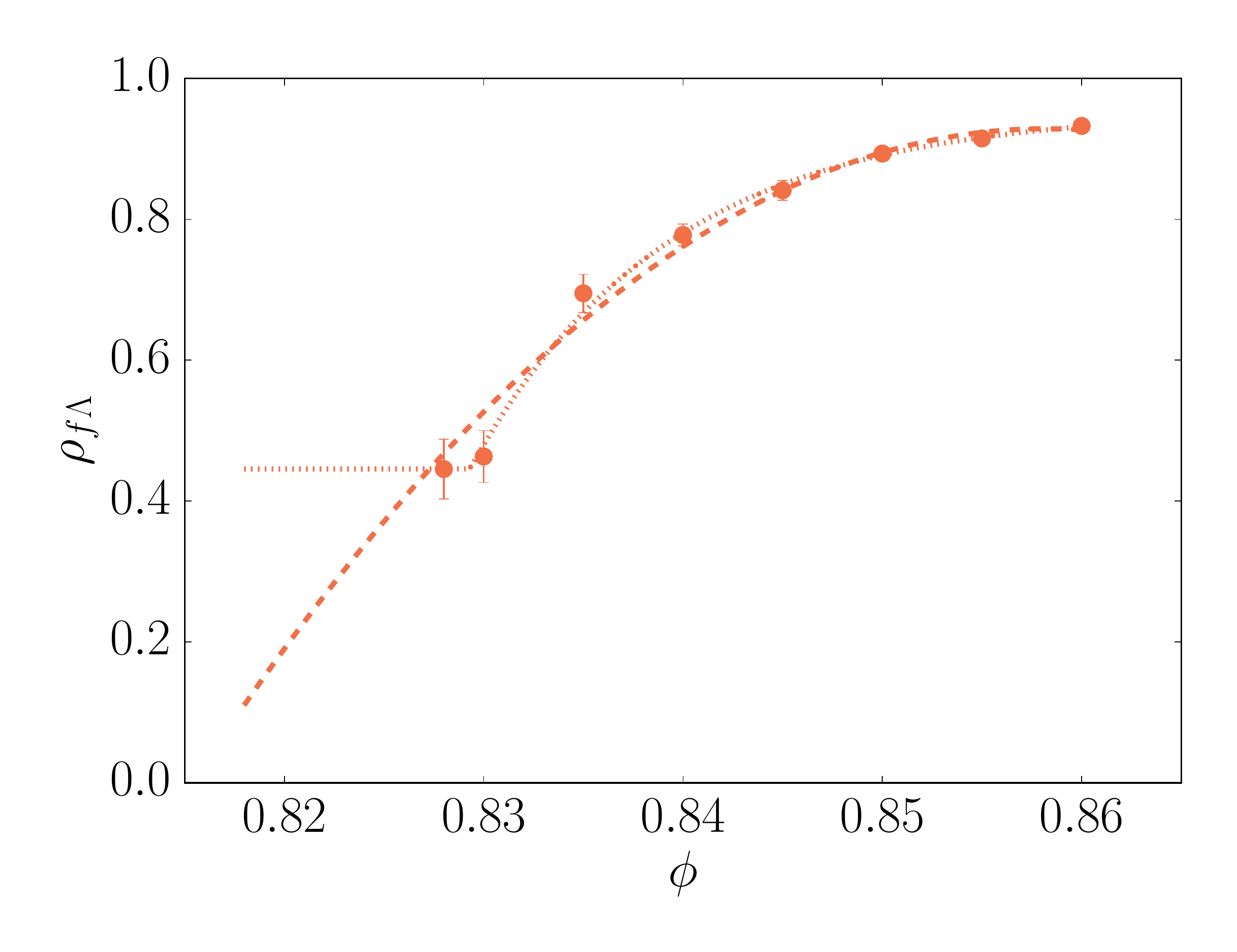}
	\caption{}
\end{subfigure}
\caption{\label{fig::biased_pdf_bv_moments} Moments of the joint
  distribution $\mathcal{B}(f, \Lambda)$ for jammed packings of $N=64$
  HS-WCA polydisperse disks at various volume fractions $0.828 \leq
  \phi \leq 0.86$. Elements of the mean $\vect{\mu} = (\mu_f,
  \mu_\Lambda)$ are shown in (a) and (b) respectively. Elements of the
  covariance matrix $\hat{\vect{\sigma}} = ((\sigma_{f}^2,
  \sigma_{f\Lambda}^2),(\sigma_{f\Lambda}^2, \sigma_{\Lambda}^2))$ are
  shown in (c)-(e). The linear correlation coefficient
  $\rho_{f\Lambda}=\sigma^2_{f\Lambda}/(\sigma_f\sigma_\Lambda)$ is
  shown in (f). All values are computed by the MCD estimator with $0.99$ support fraction over 1000 bootstrap samples. Error bars are standard errors computed by bootstrap. Dashed lines are second order polynomial fits and dotted lines are sigmoid fits (Eq.~\ref{eq::sigmoid}). Curves of best fit are meant as guide to the eye.}
\end{figure}

 In Fig.~\ref{fig::biased_pdf_bv} we show the biased distributions
 $\mathcal{B}(F)$ and $\mathcal{B}(\Lambda)$ of the basin negative log-volumes
 and log-pressures, which are the marginal distributions of the joint
 distribution $\mathcal{B}(F,\Lambda)$ shown in Fig.~2b of the main text. In
 Fig.~\ref{fig::biased_pdf_bv_moments} we plot the moments of
 $\mathcal{B}(F,\Lambda)$, namely the elements of the mean $\vect{\mu}
 = (\mu_f, \mu_\Lambda)$ and the elements of the covariance matrix
 $\hat{\vect{\sigma}} = ((\sigma_{f}^2,
 \sigma_{f\Lambda}^2),(\sigma_{f\Lambda}^2, \sigma_{\Lambda}^2))$, as
 well as the linear correlation coefficient $\rho_{f\Lambda} =
 \sigma^2_{f\Lambda} / (\sigma_f \sigma_\Lambda)$.

\section{Estimates of the equiprobability density $\phi^*_{N=64}$}

We summarise the estimated values for $\phi^*_{N=64}$ in
Table~\ref{tab::phistar}

\begin{table}[h]
\centering
\begin{tabular}{c*{3}{c}}
~              & $\lambda$ & $S_B^{(Gauss)} $ & $S_B^{(KDE)}$ \\
\hline
$\phi^*_{N=64}$ & $0.82 \pm 0.07$ & $0.82$ & $0.82$  \\
$\langle z \rangle_{\text{sig}}(\phi^*_{N=64})$ & $4.1 \pm 0.2$ & $4.0$ & $4.0$  \\
\end{tabular}
\caption{\label{tab::phistar} Predicted values of $\phi^*_{N=64}$
  obtained from the linear extrapolation of $\lambda \rightarrow 0$
  and from the point of intersection of the Gibbs entropy $S_G$ with
  the Boltzmann entropy $S_B$, computed both parametrically by fitting
  $\mathcal{B}(f)$ with a generalised Gaussian function (`Gauss') and
  non-parametrically by computing a Kernel Density Estimate (`KDE') of
  the distribution. The corresponding average contact number has been
  computed using a sigmoid fit (Eq.~\ref{eq::sigmoid}) of the data in
  Fig.~\ref{fig::phi_meanZ }.}
\end{table}

\section{Contact number}

\begin{figure}
    \includegraphics[width=0.65\textwidth]{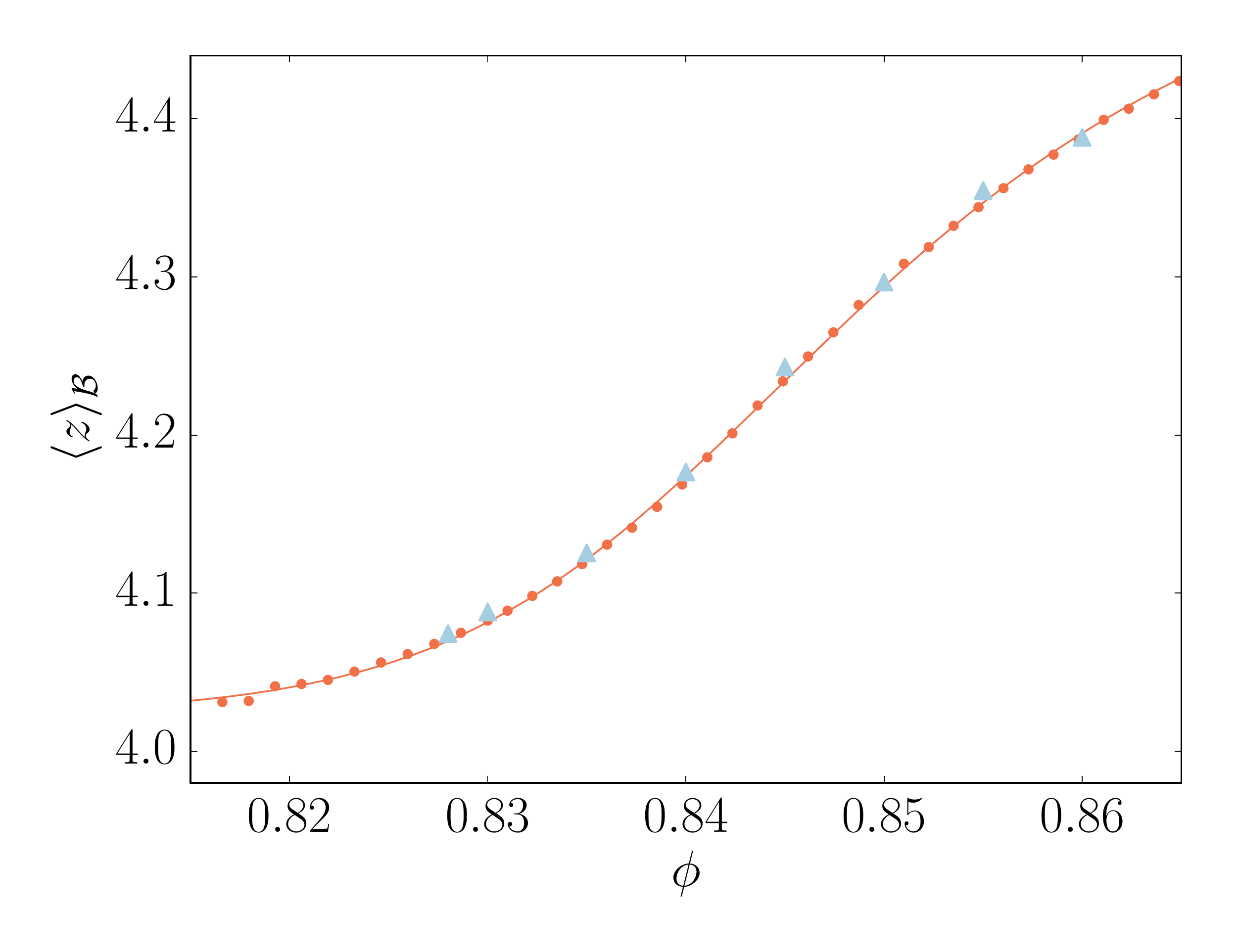}
    \caption{\label{fig::phi_meanZ } Observed average contact number
      for jammed packings of $N=64$ HS-WCA polydisperse
      disks. Triangles are estimates from the basin volume
      measurements datasets. Circles are estimates from the
      independent measurements used for the finite size scaling
      analysis. The solid line is a generalised sigmoid fit (Eq.~\ref{eq::sigmoid}) of the latter.}
\end{figure}

\begin{figure}
    \includegraphics[width=0.65\columnwidth]{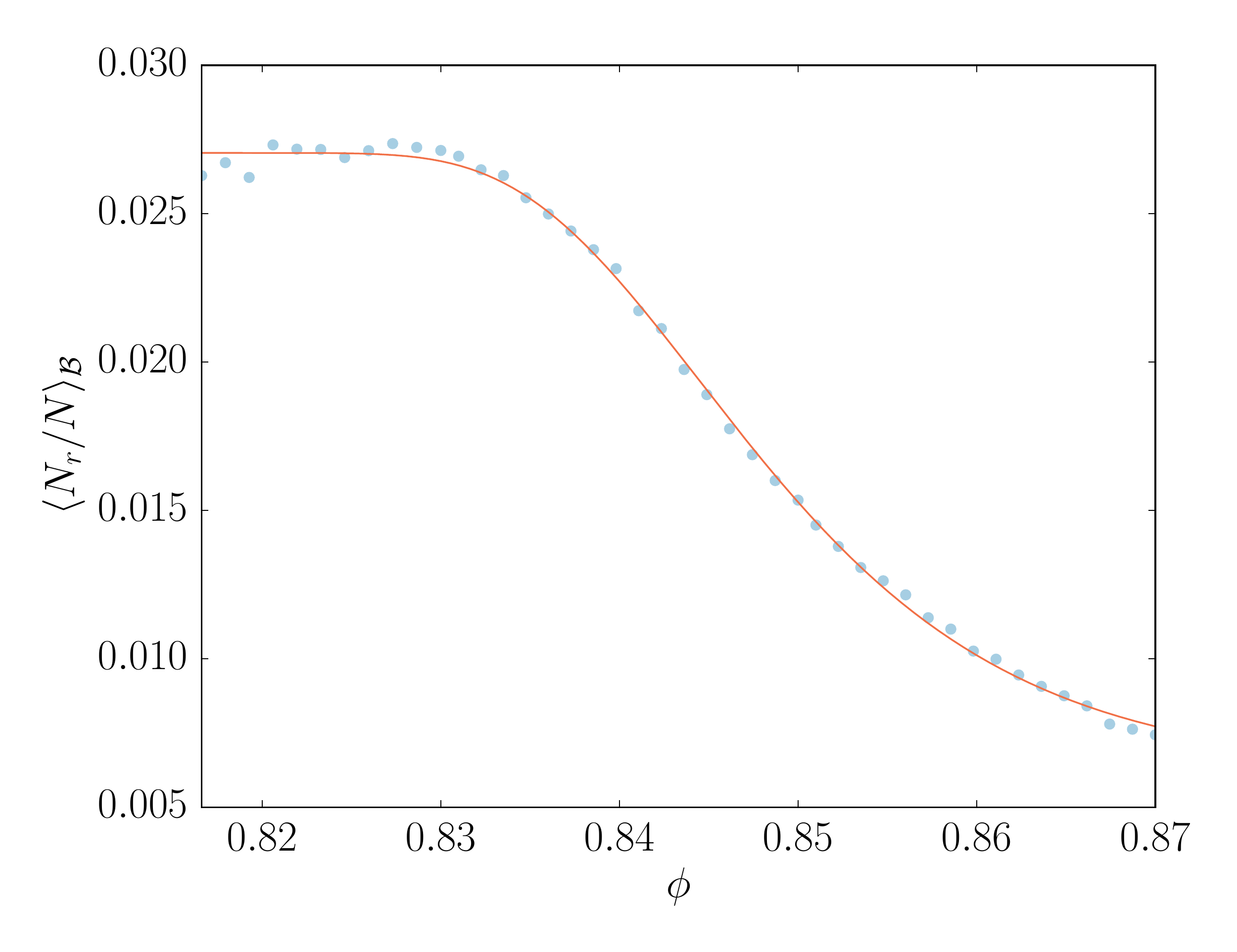}
    \caption{\label{fig::rattlers_density} Fraction of rattlers for
      jammed packings of $N=64$ HS-WCA polydisperse disks. We find
      this tends to a maximum and saturates over the same range of
      packing fractions where relative pressure fluctuations are
      maximal, see Fig.~3 of the main text. The solid line is a generalised sigmoid
      fit, Eq.~\ref{eq::sigmoid}.}
\end{figure}

The mean contact number is plotted as a function of volume fraction in
Fig.~\ref{fig::phi_meanZ }. The data are fitted with a generalised
sigmoid function of the form
\begin{equation}
\label{eq::sigmoid}
f(a, b, \phi_0, u, w; \phi) = a - \frac{a-b}{(1+e^{-w(\phi-\phi_0)})^{1/u}}
\end{equation}
In Fig.~\ref{fig::rattlers_density} we also show the fraction of
rattlers at different packing fractions for $N=64$ disks. We note that
the fraction of rattlers is maximal and saturates over the same range
of densities as where the relative pressure fluctuations are maximal
and saturate (see Fig.~3 of the main text and the finite size scaling section for further
discussion).

\clearpage

\section{Correlations with structural parameters}

We analyse the correlation of the basin negative log-volume with a number of structural parameters other than the pressure $P = \mathrm{Tr}(\hat{\bf \Sigma})/(dL^d)$, where $\hat{\bf \Sigma}$ is the stress tensor and $d=2$, discussed in detail in the main text (see Fig.~2). 

For all observables $X$ we assume a linear correlation defined analogously to Eq.~1, namely 

\begin{equation}
\label{eq::struct_linear_fit}
\langle f \rangle_{\mathcal{B}}(\phi ; X) = \lambda_X(\phi) \ln X + c_X(\phi).
\end{equation} 

\begin{figure}
\begin{subfigure}{0.5\textwidth}
  \centering
    \includegraphics[width=\linewidth]{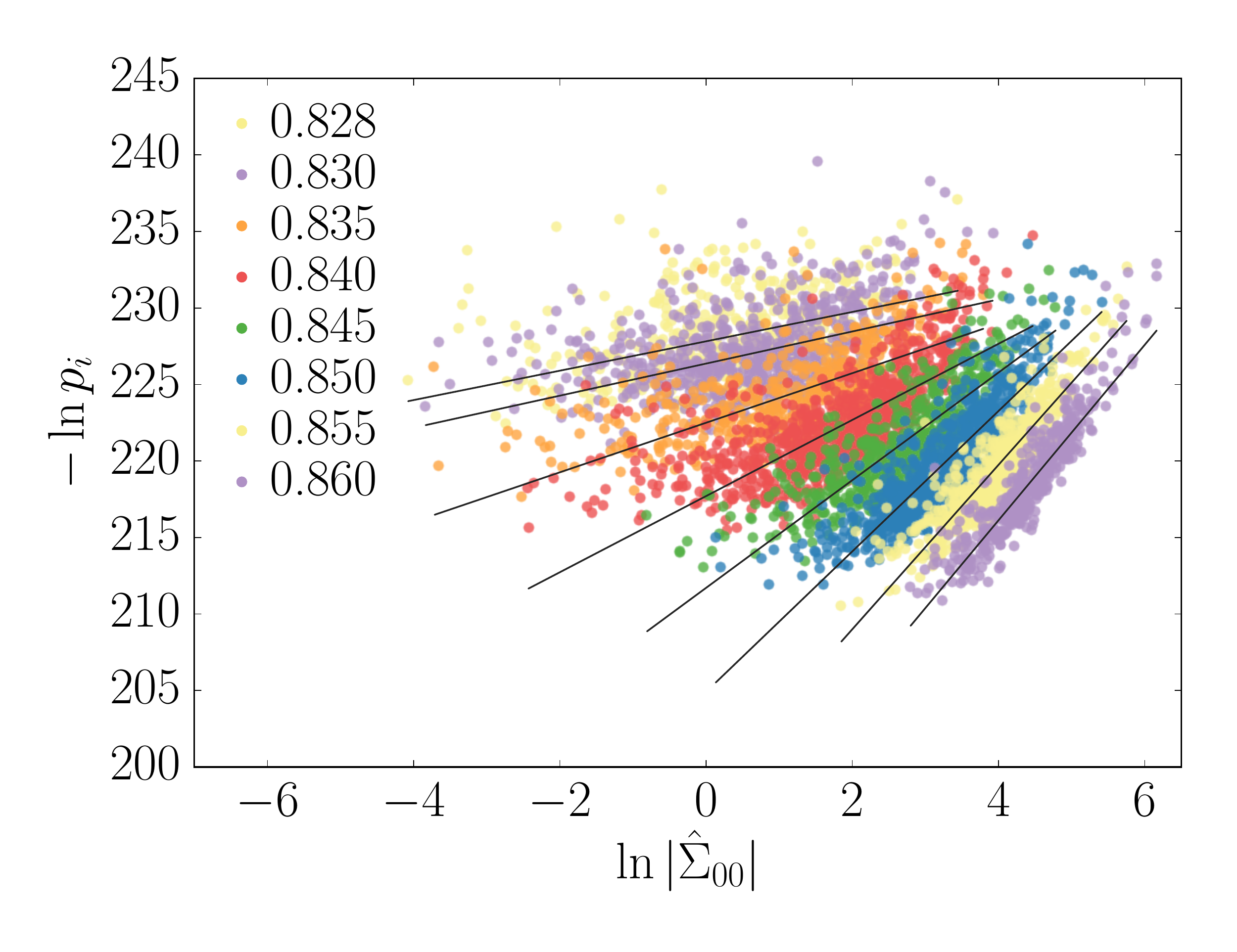}
	\caption{}
\end{subfigure}%
\begin{subfigure}{0.5\textwidth}
  \centering
    \includegraphics[width=\linewidth]{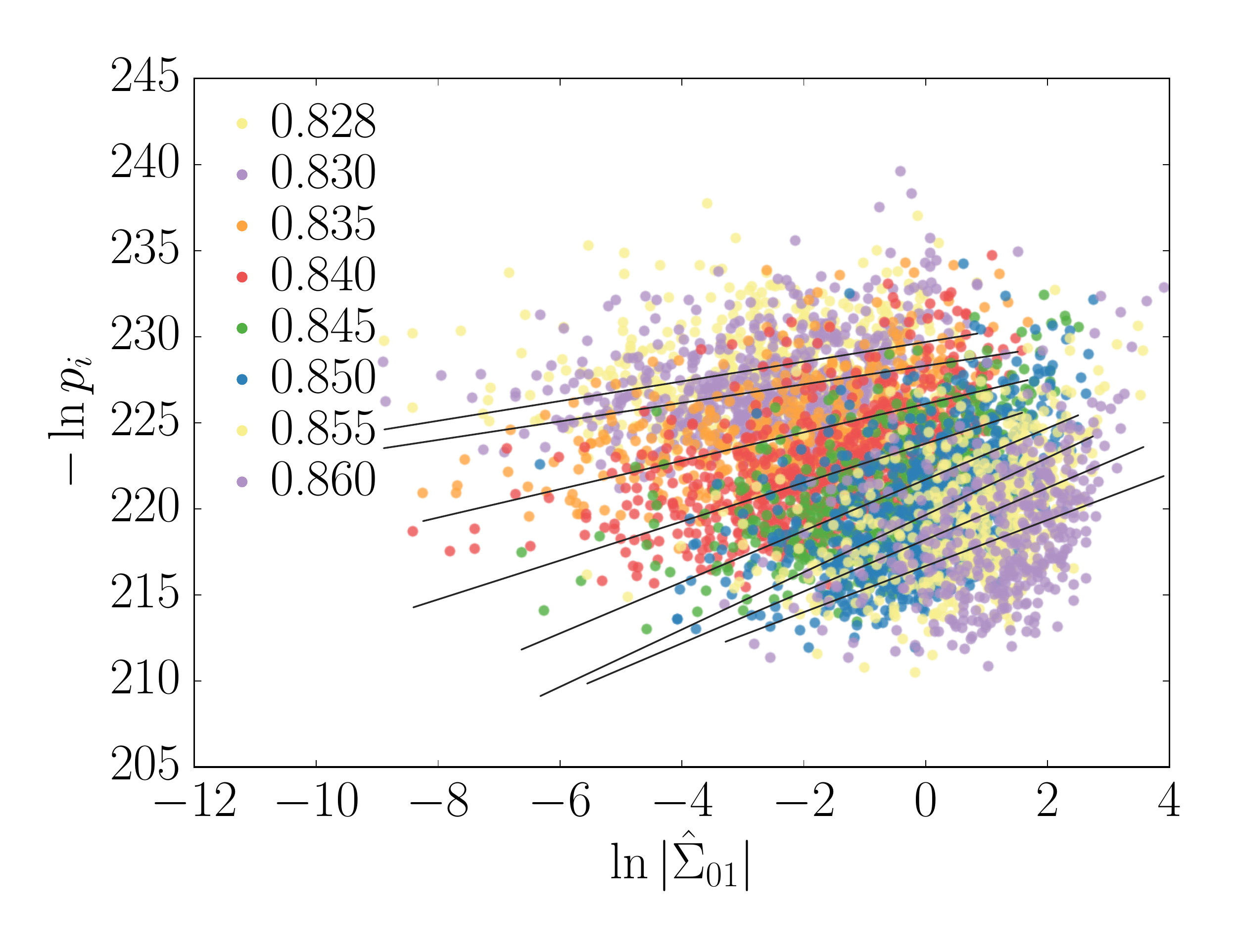}
	\caption{}
\end{subfigure}
\begin{subfigure}{0.5\textwidth}
  \centering
    \includegraphics[width=\linewidth]{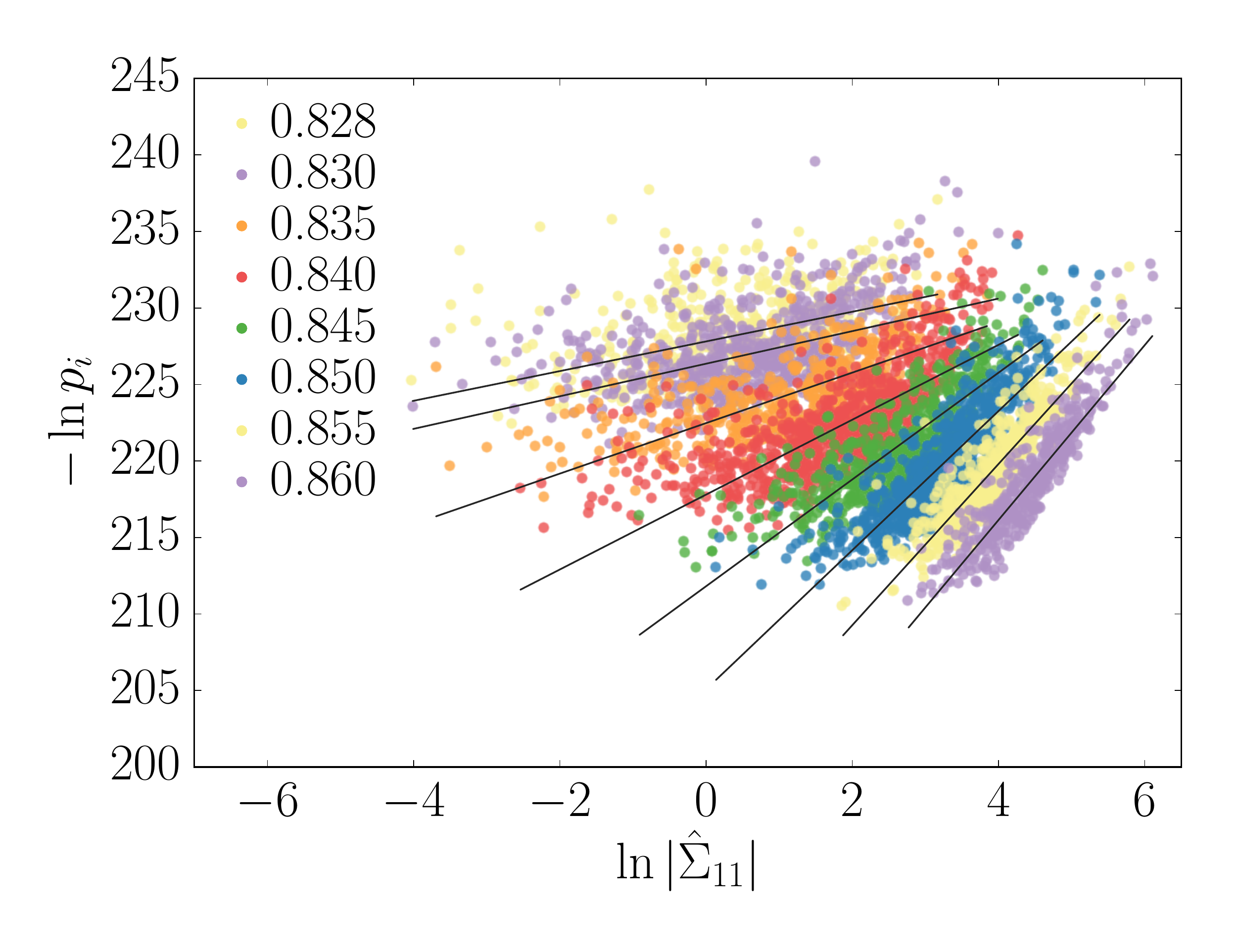}
	\caption{}
\end{subfigure}%
\begin{subfigure}{0.5\textwidth}
  \centering
    \includegraphics[width=\linewidth]{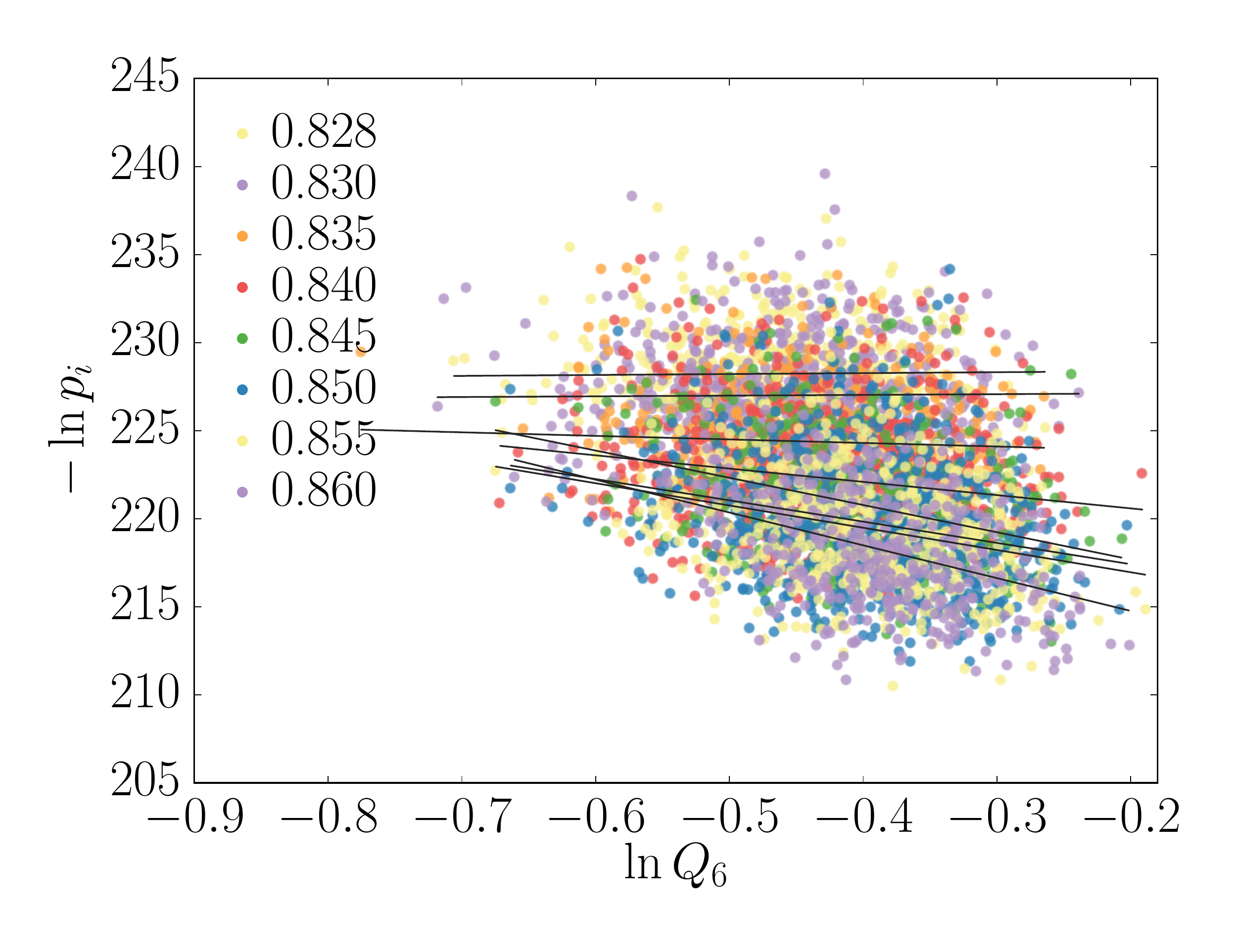}
	\caption{}
\end{subfigure}
\begin{subfigure}{0.5\textwidth}
  \centering
    \includegraphics[width=\linewidth]{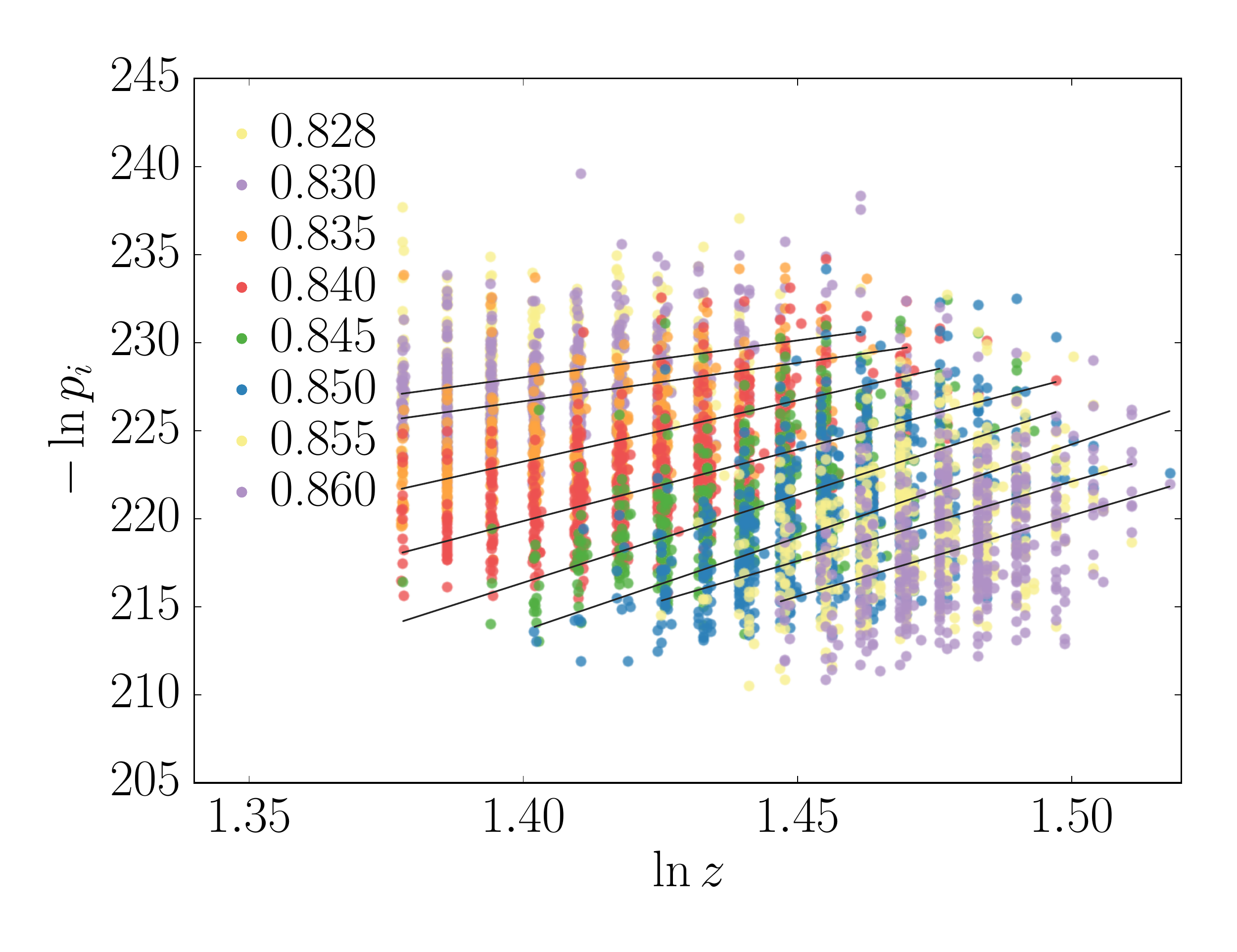}
	\caption{}
\end{subfigure}%
\begin{subfigure}{0.5\textwidth}
  \centering
    \includegraphics[width=\linewidth]{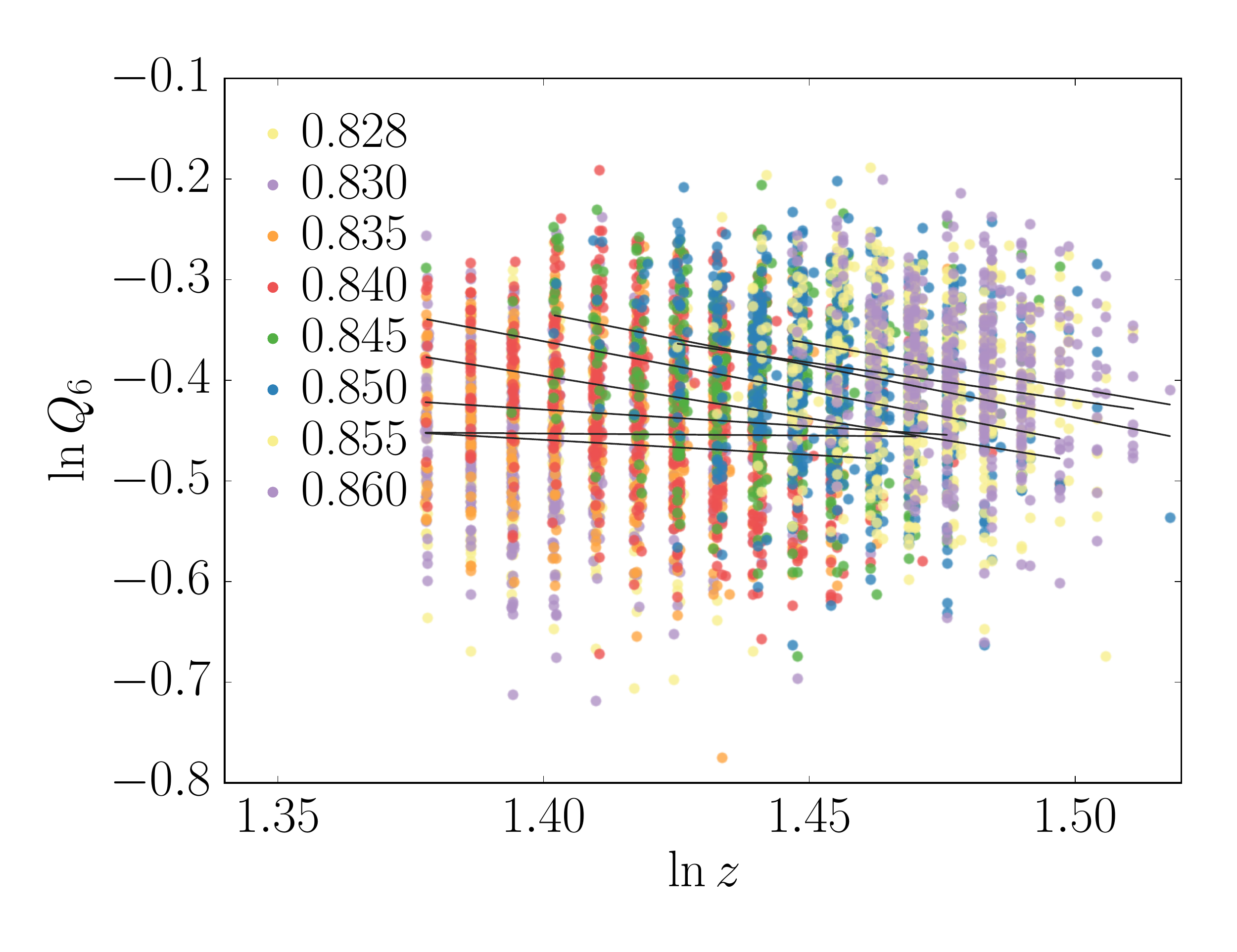}
	\caption{}
\end{subfigure}
\caption{\label{fig::struct_correlations} Scatter plots of the negative log-probability of observing a packing, $-\ln p_i = F_i + \ln V_J(\phi)$, where $V_J$ is the accessible fraction of phase space, as a function of the individual terms of the stress tensor $\hat{\bf \Sigma}$ (a)-(c), the $Q_6$ bond-orientational order parameter (d) and the average contact number $z$ (e). The scatter plot in (f) shows the $Q_6$ bond-orientational order parameter as a function of the average contact number $z$. Black solid lines are lines of best fit computed by bootstrapped linear MMSE using a robust covariance estimator.}
\end{figure}

\begin{figure}
\begin{subfigure}{0.5\textwidth}
  \centering
    \includegraphics[width=\linewidth]{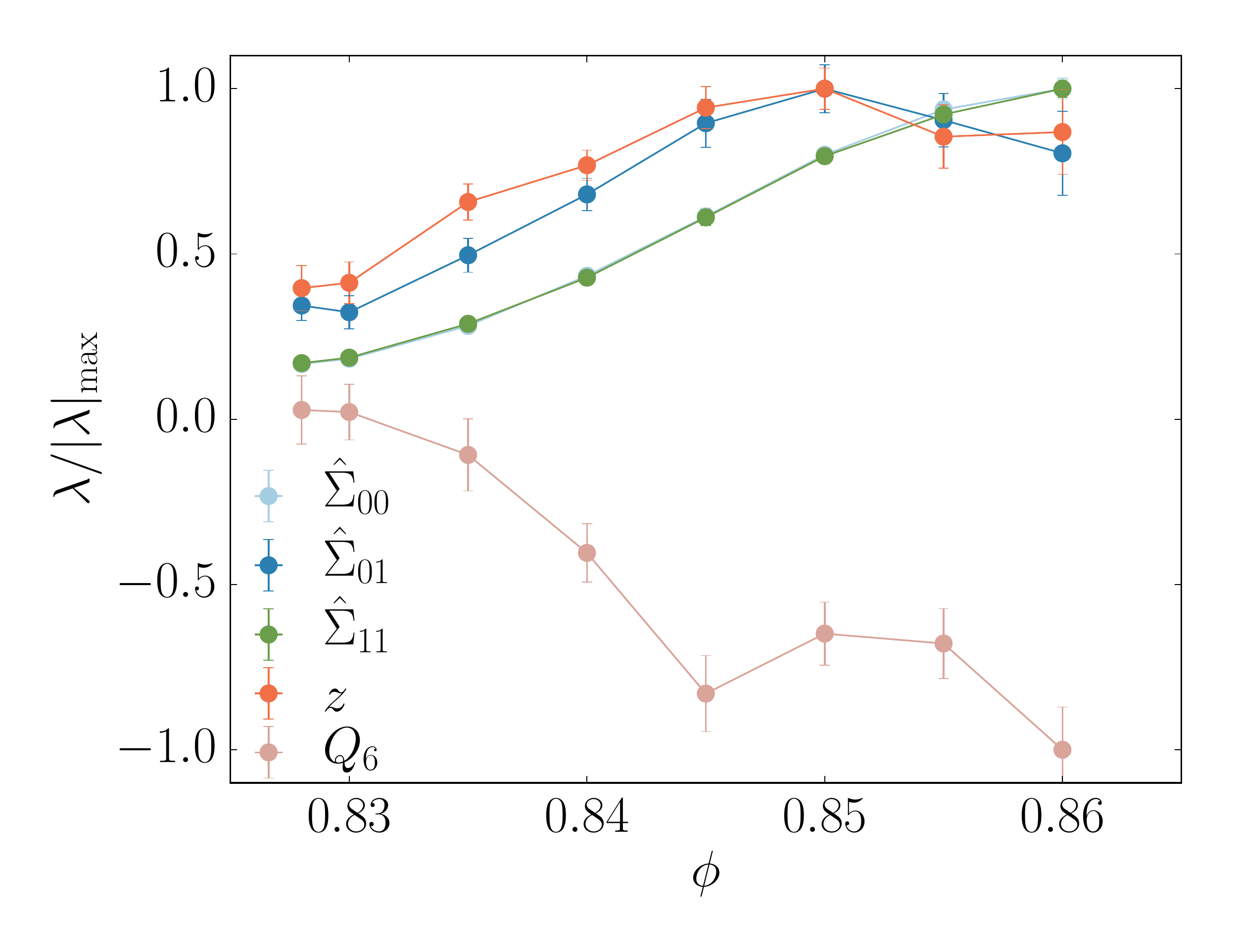}
	\caption{}
\end{subfigure}%
\begin{subfigure}{0.5\textwidth}
  \centering
    \includegraphics[width=\linewidth]{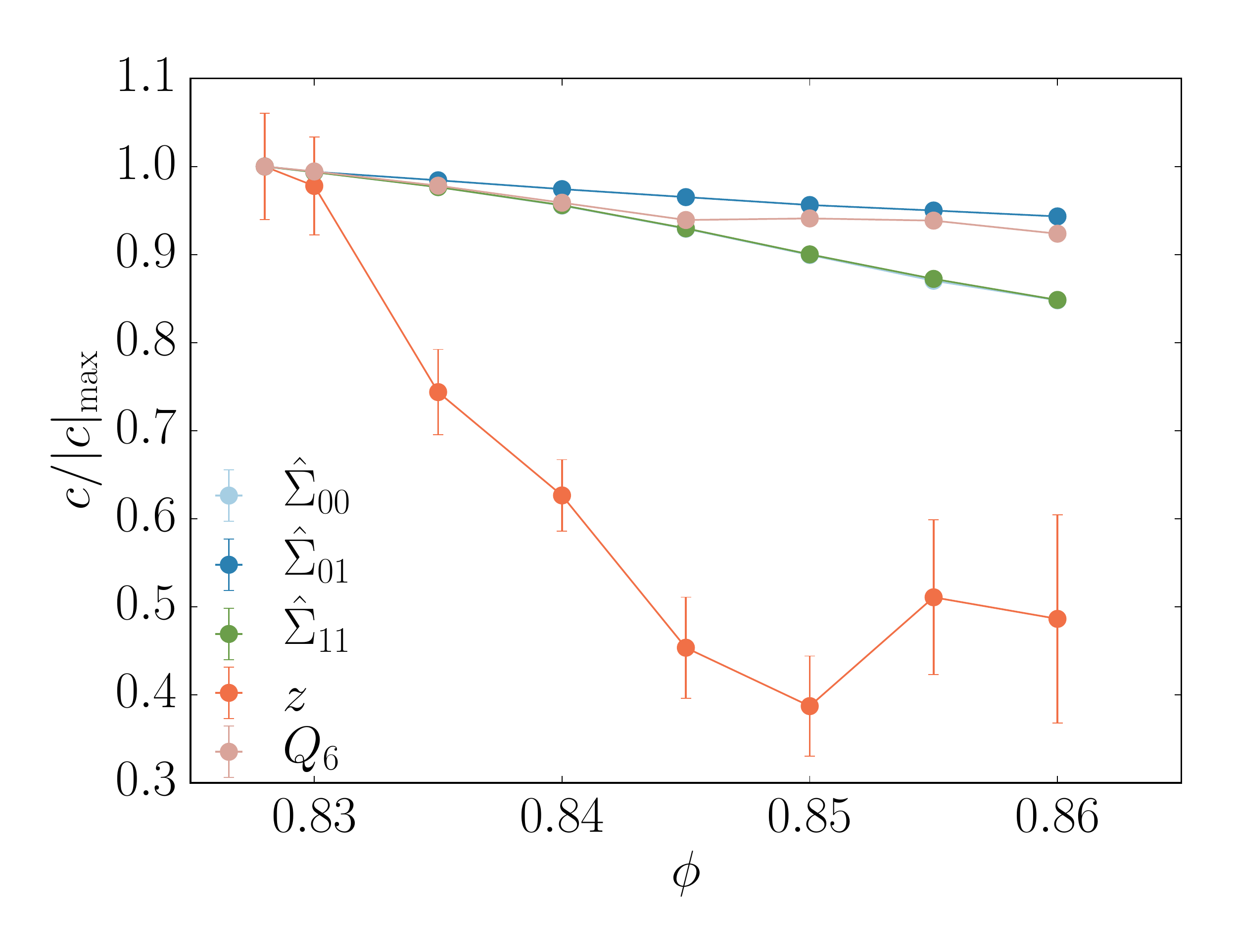}
	\caption{}
\end{subfigure}
\caption{\label{fig::struct_correlations_params} Slopes $\lambda_X$ (a) and intercepts $c_X$ (b) of Eq.~\ref{eq::struct_linear_fit} for the individual components of the the stress tensor $\hat{\bf \Sigma}$, the $Q_6$ bond-orientational order parameter, and the average contact number $z$. Estimates were obtained by bootstrapped linear MMSE fits using a robust covariance estimator and error bars refers to the standard error computed by bootstrap. Solid lines are guide to the eye.}
\end{figure}

We perform the analysis for the individual elements of the stress tensor $\hat{\bf \Sigma}_{ij}$, the average contact number $z$ and the $Q_6$ bond-orientational order parameter \cite{Steinhardt83}. Scatter plots with bootstrapped linear MMSE fits are shown in Fig.~\ref{fig::struct_correlations}, and the fitted parameters are plotted as a function of volume fraction in Fig.~\ref{fig::struct_correlations_params}. The results are qualitatively similar to those obtained for the pressure in that $\lambda_X$ is decreasing towards $0$ as $\phi \to \phi^*$, indicating that the basin volumes decorrelate from X in this limit. As explained in the main text, this is a necessary condition for the equiprobability of jammed states. 

In Fig.~\ref{fig::struct_correlations}d, we observe that $\lambda_{Q_6}$ becomes precisely zero at the lowest volume fractions while for larger volume fractions $\lambda_{Q_6}<0$, implying that larger basins correspond to more ordered structures. At the same time we note from Figs.~\ref{fig::struct_correlations}e-f that larger volumes correspond on average to lower average contact numbers ($z$) and that $z$ and $Q_6$ are (therefore) negatively correlated.  

\section{Global model of $\mathcal{B}(F, \Lambda)$}

\begin{figure}
\begin{subfigure}{0.5\textwidth}
  \centering
    \includegraphics[width=\linewidth]{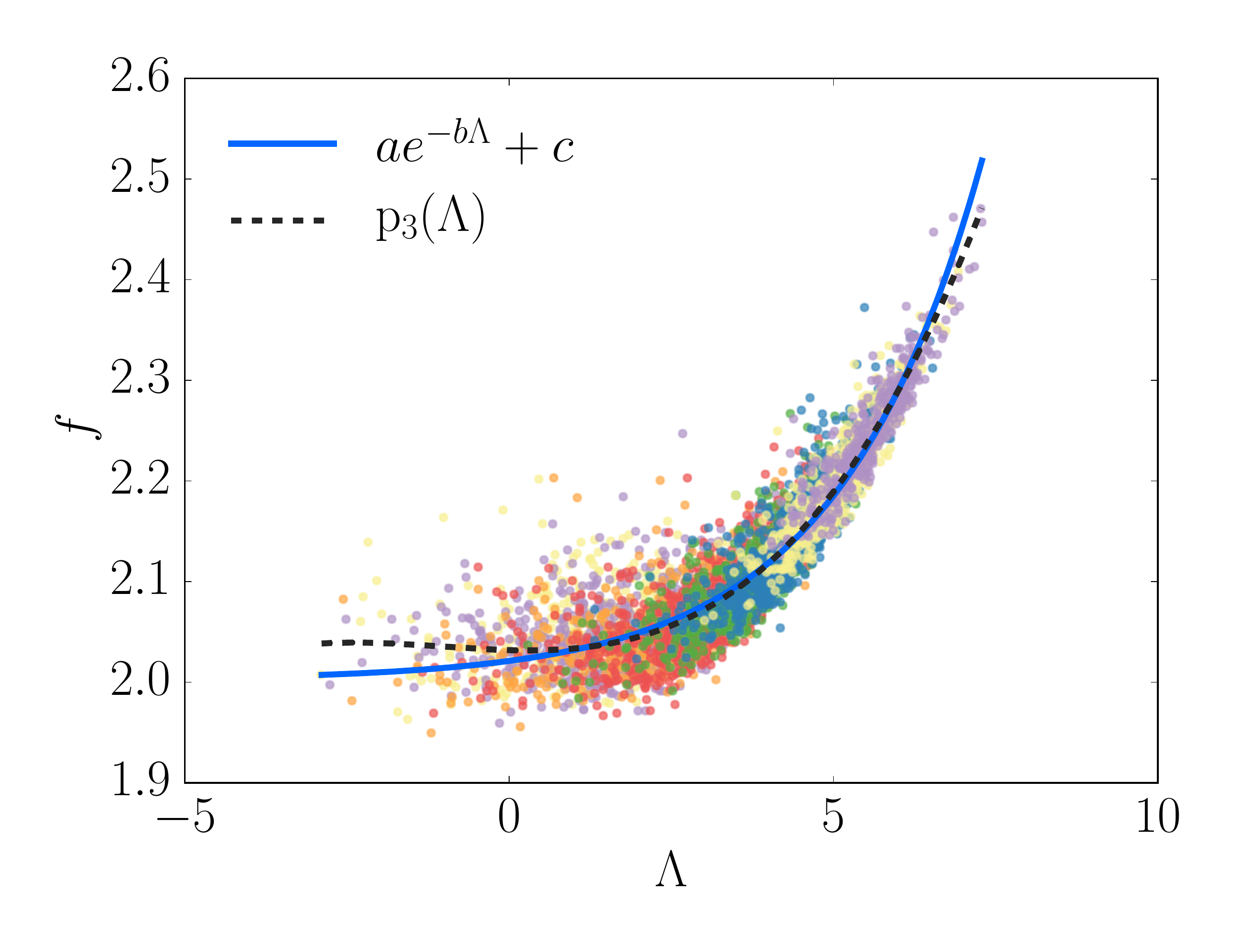}
	\caption{}
\end{subfigure}%
\begin{subfigure}{0.5\textwidth}
  \centering
    \includegraphics[width=\linewidth]{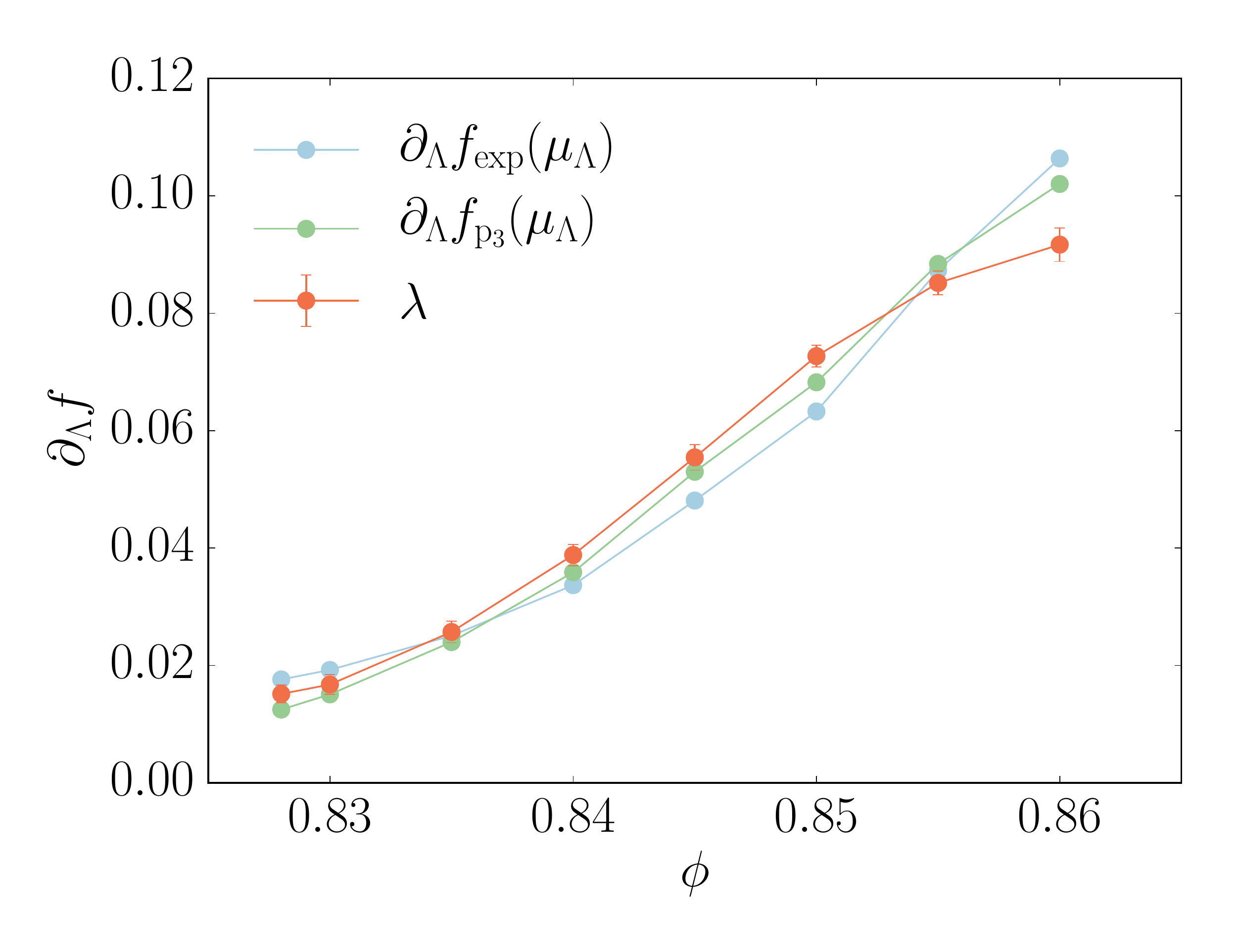}
	\caption{}
\end{subfigure}
\caption{\label{fig::global_fit} (a) Global fit of $\mathcal{B}(f,\Lambda)$ by an exponential function of the form $a\exp(-b\lambda) + c$, and a third order polynomial $\text{p}_3(\Lambda)$. (b) First derivative of the fits evaluated at mean log-pressure $\mu_\Lambda(\phi)$ are in excellent agreement with the estimates of $\lambda$ obtained by linear MMSE, see Eq.~17. Solid lines are guide to the eye.}
\end{figure}

In Fig.~2b of the main text, we fit the joint probability $\mathcal{B}(\phi; F,\Lambda)$ by linear MMSE, or in other words we fit the data with a bivariate Guassian such that the conditional expectation of F given $\Lambda$ corresponds to the linear fit. We compute linear MMSE fits for each volume fraction $\phi$ independently, as the probabilities of the packings ($-\ln p_i = F_i + \ln V_J(\phi)$) are obtained by subtracting different normalization constants for each $\phi$ (accessible volume $V_J(\phi)$). However, since $\mu_f(\phi)  \equiv \langle F/N \rangle_{\mathcal{B}}(\phi)$ and $\mu_\Lambda \equiv \langle \Lambda \rangle_{\mathcal{B}}(\phi)$ are slowly varying (see Fig.~\ref{fig::biased_pdf_bv_moments}a-b), we attempt to fit to the full distribution of $\mathcal{B}(f,\Lambda)$ (for all $\phi$ at once) by an exponential function of the form $a\exp(-b\lambda) + c$, and a third order polynomial $\text{p}_3(\Lambda)$. Fits are shown in Fig.~\ref{fig::global_fit}a, showing an evident decay of the correlations between pressure and basin volume. Evaluating the derivative of these global fits at each $\mu_\Lambda(\phi)$ we find that they are in excellent agreement with the estimates of $\lambda$ obtained by linear MMSE, see Eq.~17 (Methods).

\section{Finite size scaling}

\begin{figure}
\begin{subfigure}{0.5\textwidth}
  \centering
    \includegraphics[width=\linewidth]{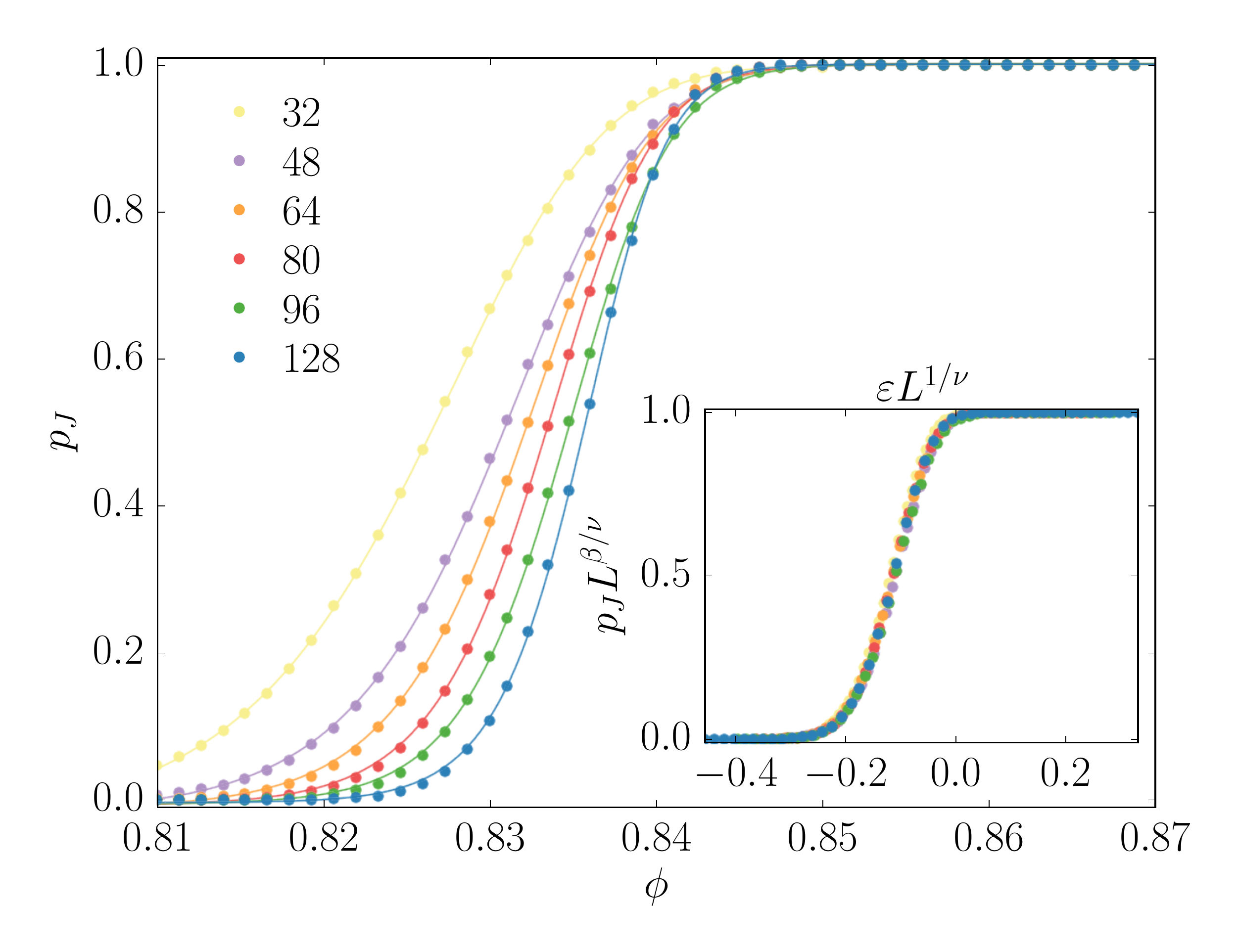}
	\caption{}
\end{subfigure}%
\begin{subfigure}{0.5\textwidth}
  \centering
    \includegraphics[width=\linewidth]{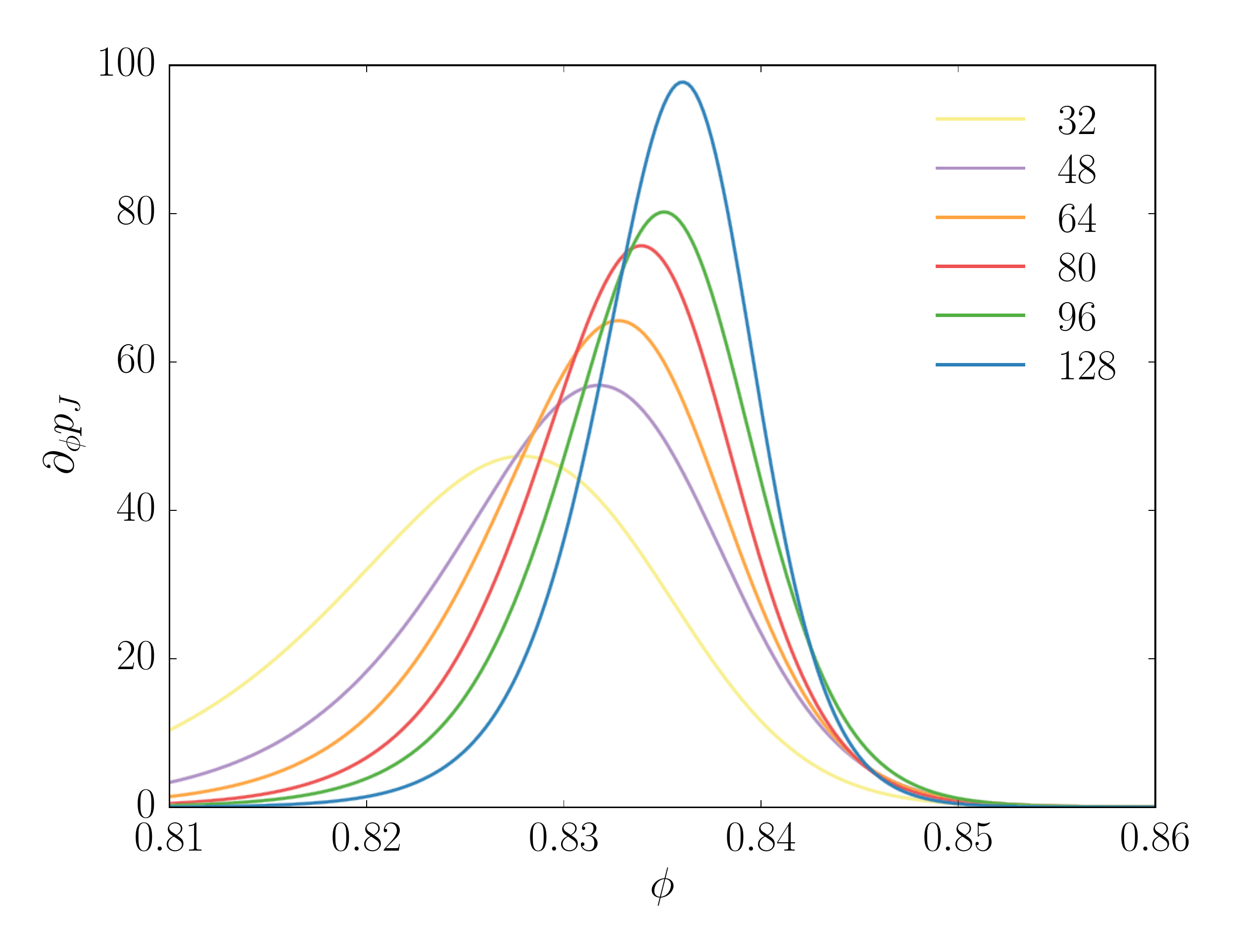}
	\caption{}
\end{subfigure}
\caption{\label{fig::phi_ppack} (a) Probability of obtaining a jammed
  packing $p_J$ by our preparation protocol for $N = 32$ to $128$ HS-WCA
  polydisperse disks as a function of volume fraction. Inset: Scaling
  collapse for $p_J L^{\beta/\nu}$ vs. $L^{1/\nu} \left(\phi/\phi^J_{N
    \to \infty} -1\right)$, with $L = N^{1/d}$, yields critical
  exponents $\nu \approx 1$, $\beta=0$ and critical volume fraction
  $\phi^{J_{(p_J)}}_{N\to\infty} = 0.844(2)$. Circles are observed data and solid lines
  correspond to sigmoid fits, Eq.~\ref{eq::sigmoid}. (b) Derivative of the
  sigmoid fits for $p_J$ for different numbers of disks.}
\end{figure}

\begin{figure}
    \includegraphics[width=0.65\linewidth]{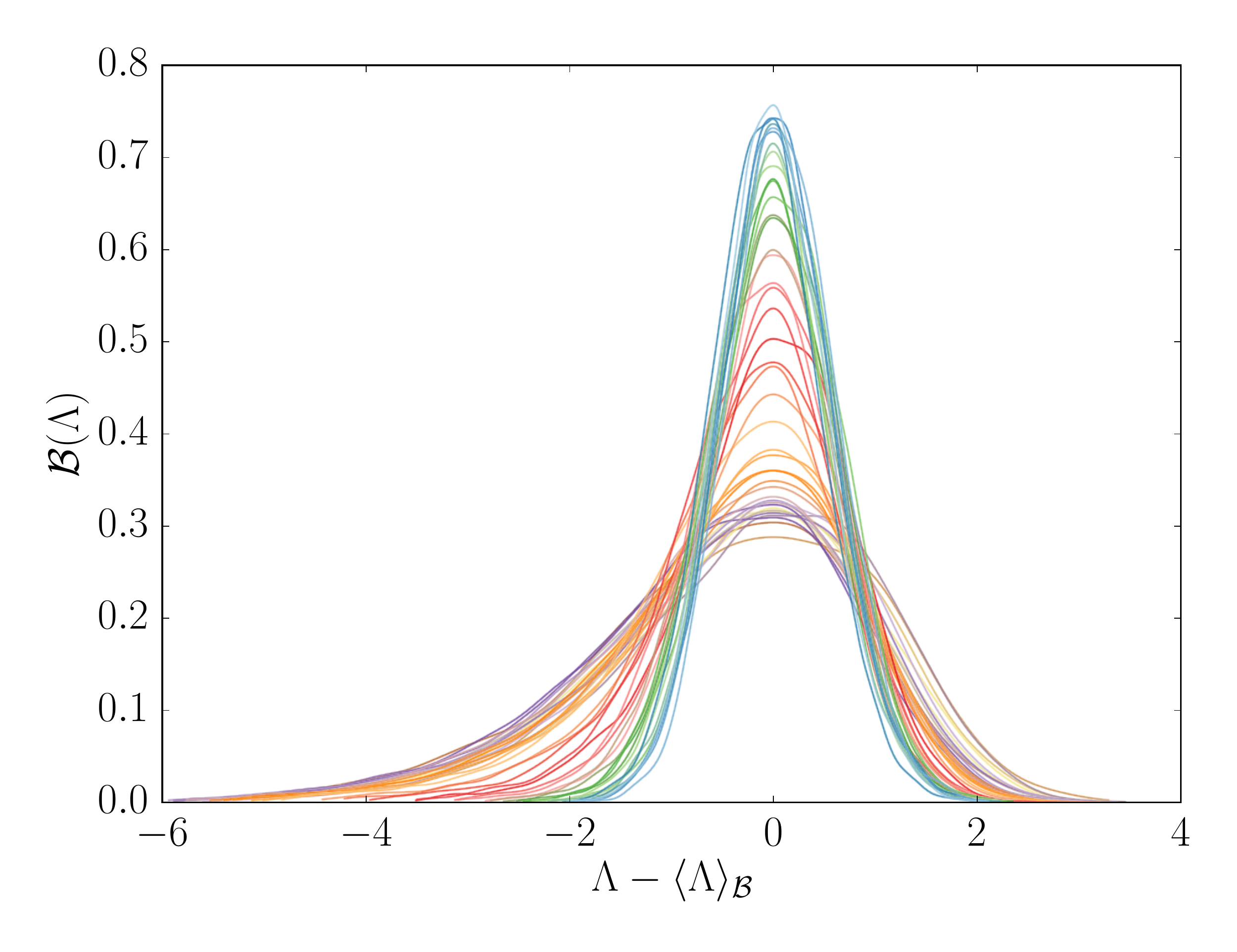}
    \caption{\label{fig::biased_pdfP} Log-transformed observed
      (biased) distribution of pressures for jammed packings of $N=64$
      HS-WCA polydisperse disks, centred around the mean. The variance
      grows for decreasing volume fractions and becomes more skewed
      towards low pressures. The overall Gaussian shape is consistent
      with a log-normal distribution of pressures. Curves are kernel density estimates with Gaussian kernels \cite{Bishop09, Scikit-learn} and bandwidth selection performed by cross validation \cite{Bowman84, Martiniani16}}
\end{figure}

\begin{figure}
\begin{subfigure}{0.5\textwidth}
  \centering
    \includegraphics[width=\linewidth]{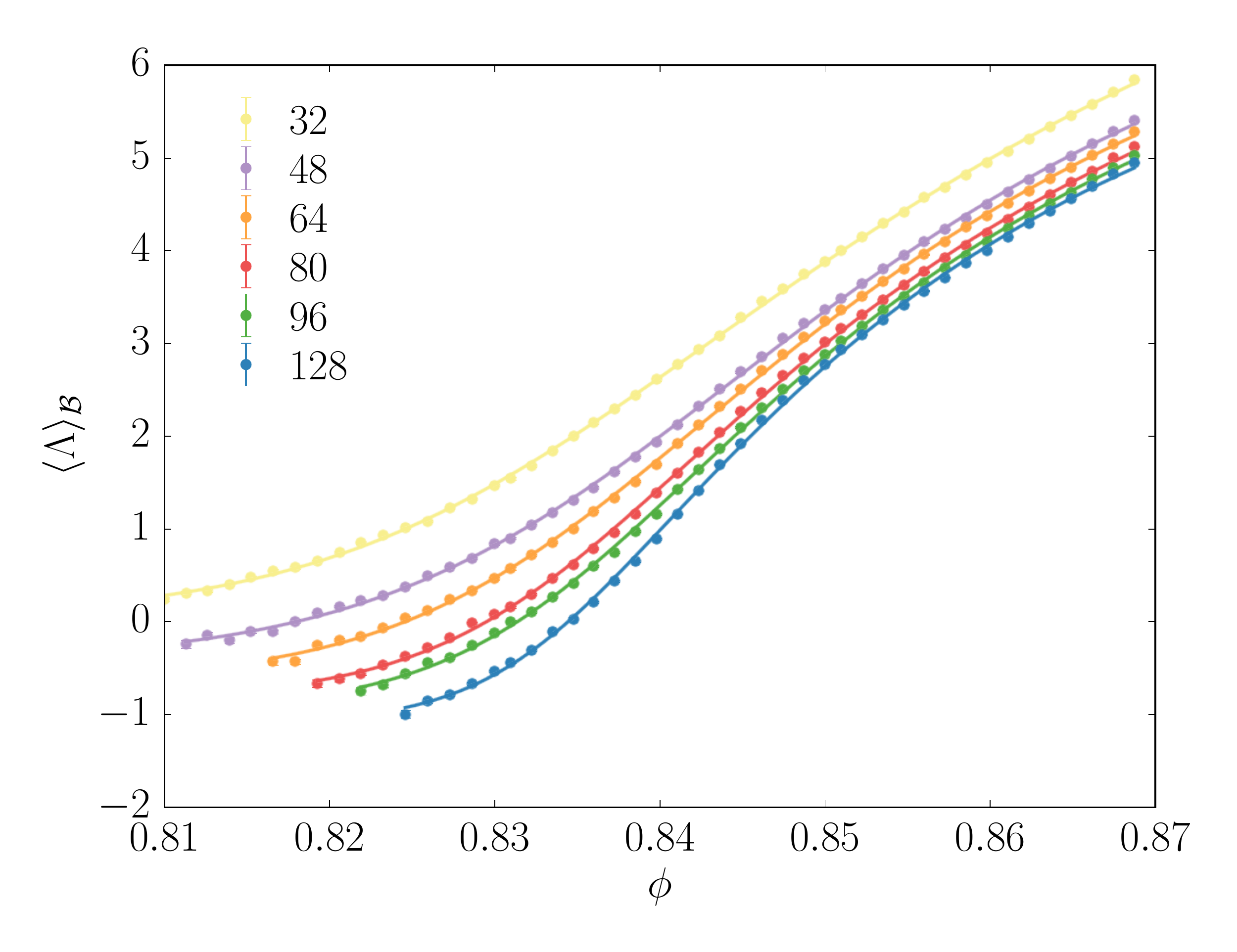}
	\caption{}
\end{subfigure}%
\begin{subfigure}{0.5\textwidth}
  \centering
    \includegraphics[width=\linewidth]{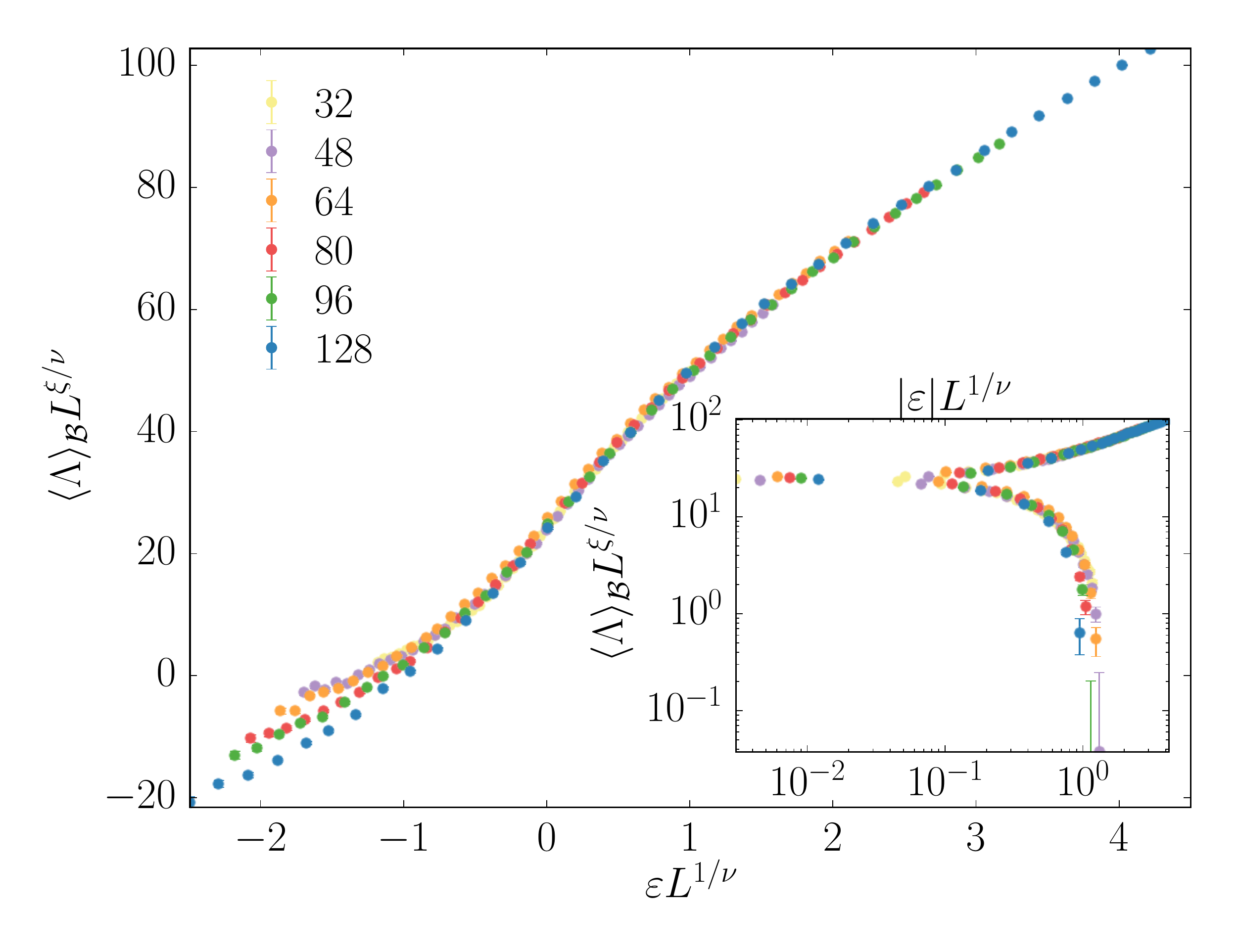}	
	\caption{}
\end{subfigure}
\caption{\label{fig::mean_lnp} (a) Average log-pressure $\langle
  \Lambda \rangle_\mathcal{B}$ for $N$ HS-WCA polydisperse disks. (b)
  Scaling collapse for $\langle \Lambda \rangle_\mathcal{B}
  L^{\xi/\nu}$ vs. $L^{1/\nu} \left(\phi/\phi^J_{N \to \infty}
  -1\right)$, with $L = N^{1/d}$. The estimated critical exponents are
  $\nu = 0.50(5)$ and $\xi = 0.62(3)$, and the critical volume
  fraction $\phi^{J_{(\Lambda)}}_{N\rightarrow\infty}=0.841(3)$.  Inset: A logarithmic
  plot of the same data. Circles are observed data and solid lines are sigmoid fits,
  Eq.~\ref{eq::sigmoid}. Error bars, computed by BCa bootstrap
  \cite{Efron87}, refer to $1\sigma$ confidence intervals.}
\end{figure}

In order to locate the unjamming transition, we compute the
probability of obtaining jammed packings as a function of volume
fraction $\phi$. A finite size scaling collapsSI for $p_J L^{\beta/\nu}$ vs. $L^{1/\nu} \left(\phi/\phi^J_{N \to \infty} -1\right)$, shown in Fig.~\ref{fig::phi_ppack}, yields critical exponents $\nu \approx 1$,
$\beta=0$ and critical volume fraction $\phi^{J_{(p_J)}}_{N\to\infty}=0.844(2)$,
in agreement with Vagberg et al. \cite{Vagberg11}. We obtain an
independent estimate of the unjamming transition by locating the point
where the average pressure goes to zero and therefore $\langle \Lambda
\rangle_\mathcal{B} \to - \infty$. KDE distributions for $\Lambda$ are
shown in Fig.~\ref{fig::biased_pdfP}. The average log-pressure is
shown in Fig.~\ref{fig::mean_lnp}a and a finite size scaling collapse for $\langle \Lambda \rangle_\mathcal{B}L^{\xi/\nu}$ vs. $L^{1/\nu} \left(\phi/\phi^J_{N \to \infty} -1\right)$, shown in Fig.~\ref{fig::mean_lnp}b, yields $\nu = 0.50(5)$, $\xi=0.62(3)$ and critical volume fraction $\phi^{J_{(\Lambda)}}_{N\to\infty}=0.841(3)$.

\begin{figure}
\begin{subfigure}{0.5\textwidth}
  \centering
    \includegraphics[width=\linewidth]{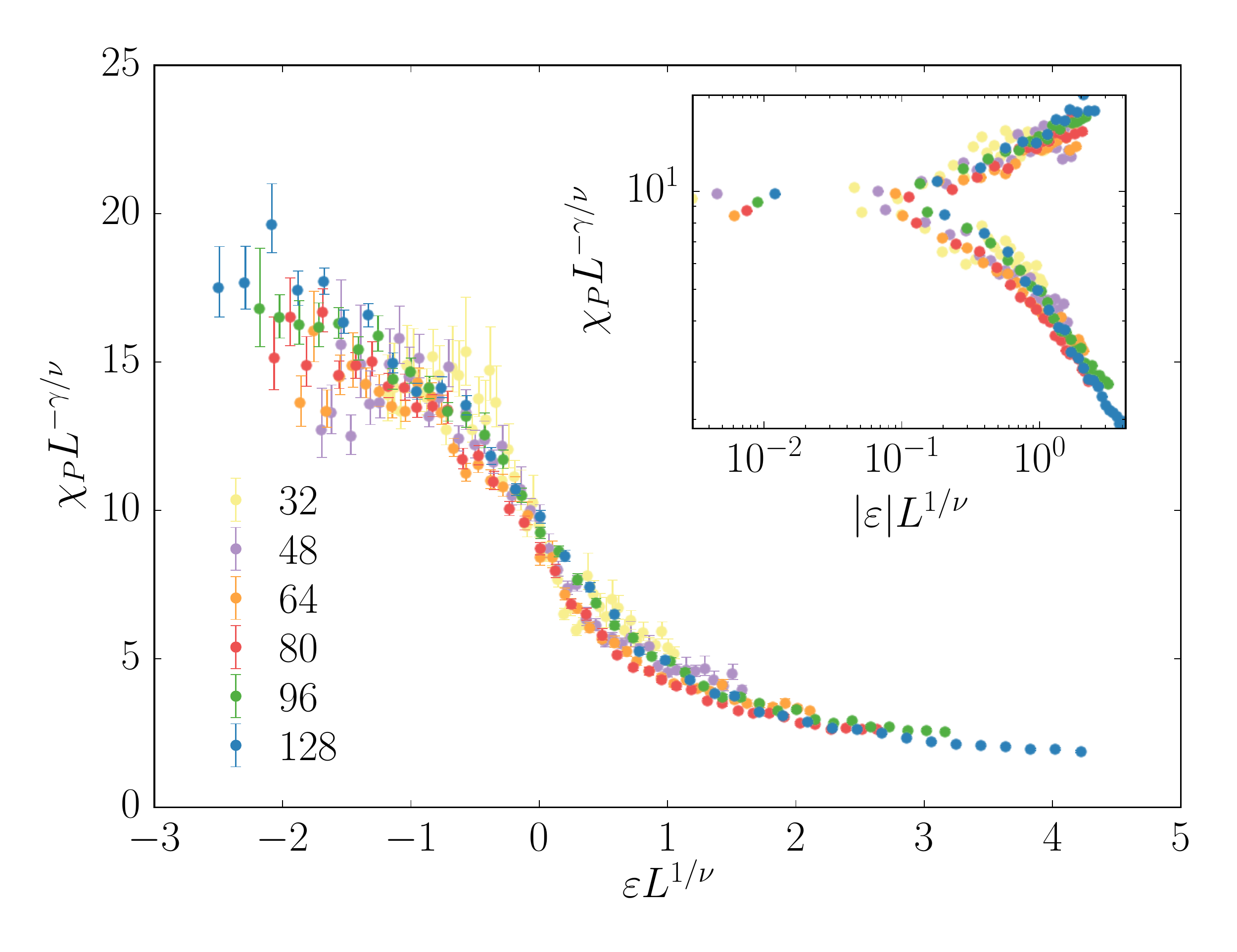}
	\caption{}
\end{subfigure}%
\begin{subfigure}{0.5\textwidth}
  \centering
    \includegraphics[width=\linewidth]{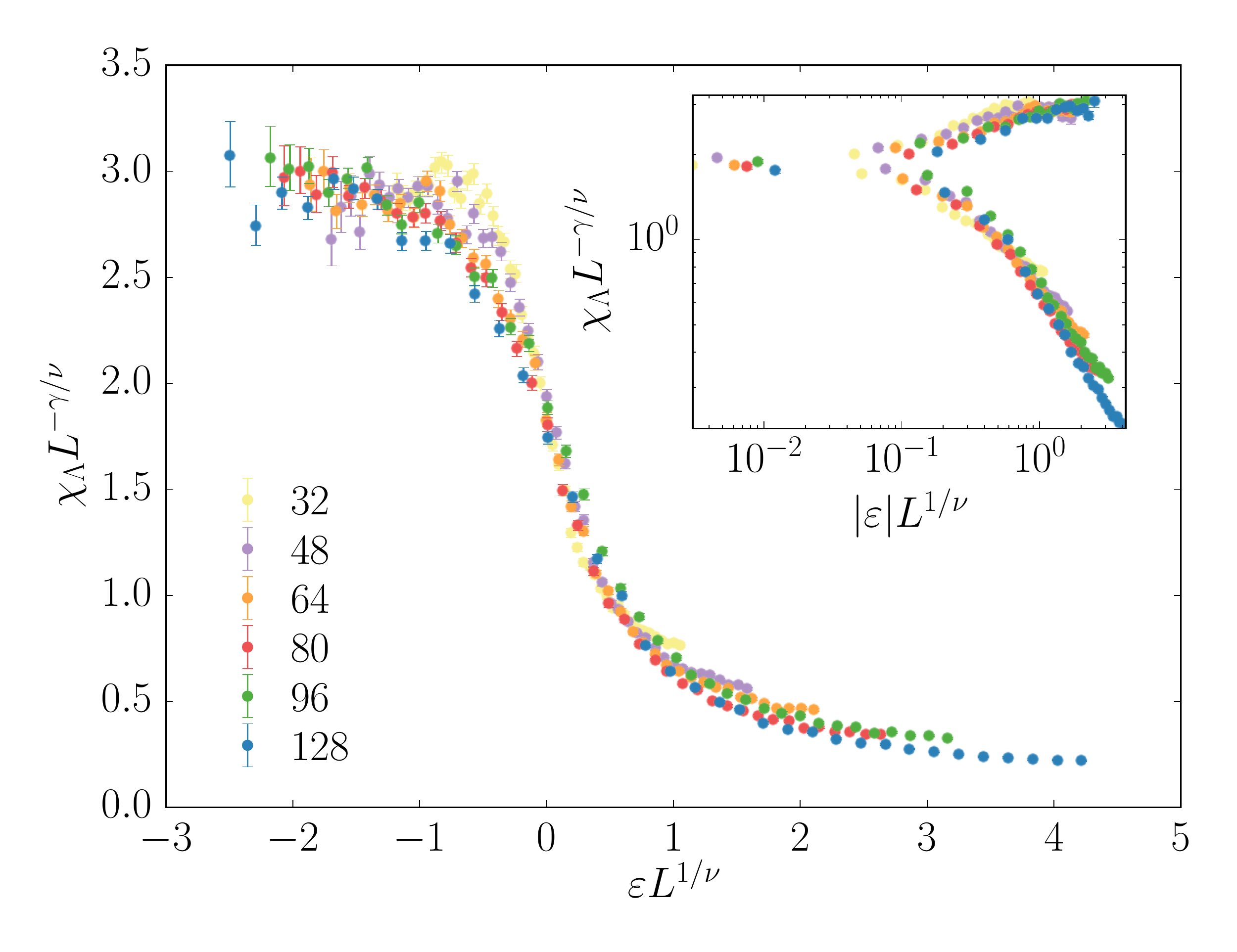}
	\caption{}
\end{subfigure}
\caption{\label{fig::var_lnp} (a) Data collapse from finite size
  scaling analysis of the variance of the relative pressures. The plot shows $\chi_{P} L^{-\gamma/\nu}$ vs. $L^{1/\nu}
  \left(\phi/\phi^J_{N \to \infty} -1\right)$, with $L = N^{1/d}$. The
  estimated critical exponents are $\nu = 0.5(3)$ and $\gamma =
  0.47(5)$, and the critical volume fraction is
  $\phi^{*_{(P)}}_{N\rightarrow\infty}=0.841(3)$. (b) Scaling collapse of the variance of the log-pressures. The plot shows
  $\chi_{\Lambda} L^{-\gamma/\nu}$ vs. $L^{1/\nu} \left(\phi/\phi^*_{N
    \to \infty} -1\right)$. The estimated critical exponents are $\nu
  = 0.5(3)$ and $\gamma = 0.89(5)$, and the critical volume fraction
  is $\phi^{*_{(\Lambda)}}_{N\rightarrow\infty}=0.841(3)$. Error bars, computed by
  BCa bootstrap, refer to $1\sigma$ confidence intervals.}
\end{figure}

We then analyse the relative pressure fluctuations $\chi_P = N
\sigma^2(P/\langle P \rangle_{\mathcal{B}})$ and the log-pressure
fluctuations $\chi_\Lambda = N \sigma^2_\Lambda$. A scaling collapse
for different system sizes of $\chi_{P} L^{-\gamma/\nu}$
vs. $L^{1/\nu} \left(\phi/\phi^J_{N \to \infty} -1\right)$ with $L =
N^{1/d}$, shown in Fig.~\ref{fig::var_lnp}a, yields $\nu= 0.5(3)$,
$\gamma=0.47(5)$ and $\phi^{*_{(P)}}_{N \to \infty} = 0.841(3)$. An analogous
scaling collapse of $\chi_{\Lambda} L^{-\gamma/\nu}$ vs. $L^{1/\nu}
\left(\phi/\phi^*_{N \to \infty} -1\right)$, shown in
Fig.~\ref{fig::var_lnp}b, yields $\nu= 0.5(3)$, $\gamma=0.89(5)$ and
$\phi^{*_{(\Lambda)}}_{N \to \infty} = 0.841(3)$.

Together these results lead us to conclude that the
point of equiprobability $\phi^*_{N \to \infty}$ coincides with the
unjamming point $\phi^J_{N \to \infty}$, to within numerical error and finite size corrections that we do not take into account. Note that
the precise numerical value of $\nu$ varies through the literature and
has been shown to depend on the quantity being observed, and also crucially on finite size corrections to scaling
\cite{Vagberg11}. In this work we have not attempted to establish $\nu$
definitely, nor elucidate its origin with respect to the diverging
correlation length(s) that might be involved.

